# Origin of the Recirculation Flow Pattern Induced by Nanosecond Discharges and Criterion for its Development

Edouard Roger[1a], Pierre Mariotto[1], Christophe O. Laux[1]

[1] Laboratoire EM2C, CNRS, CentraleSupélec, Université Paris-Saclay, 3 rue Joliot-Curie, 91190 Gif-sur-Yvette, France

**Abstract.** Nonequilibrium plasmas generated by spark discharges induce chemical, thermal, and flow dynamic effects that are beneficial in many applications such as plasmalysis, plasma-assisted combustion, and plasma flow control. Among the flow dynamic effects generated by these discharges, vorticity is of particular interest because it enhances the mixing of the discharge products with the surrounding environment and accelerates the cooling of the kernel. This article provides a comprehensive examination of the vorticity produced by nanosecond discharges in air. Using computational fluid dynamic simulations of the blast wave resulting from energy deposition in the interelectrode volume during the nanosecond pulse, we analyze the various sources of vorticity during the post-discharge period. The non-uniform strength of the leading shock of the blast wave is found to be the primary promoter of vorticity. Other sources, such as the barocline torque, play a secondary role in explaining the overall recirculation pattern. In addition, the discharge cooling mechanisms are also investigated. The cooling regimes and their efficiencies are classified according to the discharge parameters, such as the inter-electrode gap, the initial kernel temperature, or the frequency of repetitive pulses. Finally, a physics-based, non-dimensional number $\Pi^*$, equal to the ratio of the initial kernel temperature to the ambient temperature, is introduced. A numerical analysis shows that the transition between the recirculating and non-recirculating flow regimes occurs for $\Pi^*$ on the order of 10. This criterion is validated against experimental and numerical results from the literature.

E-mail: edouard.roger@centralesupelec.fr, pierre.mariotto@centralesupelec.fr, christophe.laux@centralesupelec.fr

## I. Introduction

Nanosecond repetitively pulsed (NRP) discharges are characterized by repetitive voltage pulses with short duration (~10 ns), high amplitude (~10 kV), and frequencies within 10–100 kHz. They are commonly applied between two pin electrodes separated by 1–10 mm and immersed in a gas. These discharges heat the gas in a few nanoseconds and produce radicals and chemicals of interest for various applications. For example, they can be used to dissociate $CO_2$ into CO, which is a critical step in the synthesis of carbon-neutral fuels [1,2]. They can also assist combustion by extending the Lean Blow-Out limit [3], by stabilizing flames [4] or by improving ignition [5].

---

[a] Author to whom any correspondence should be addressed.

The pulsing strategy of NRP discharges is based on repeating the pattern illustrated in Figure 1a. This pattern can be divided into the pulse and the post-discharge phases. During the pulse phase (its duration is referred to as $\tau_{\text{pulse}}$), the high voltage applied across the electrodes creates an electric field that excites, ionizes, and dissociates the gas in the interelectrode gap. Depending on the type of nanosecond discharge, the post-pulse pressure and temperature (noted $P_k$ and $T_k$) may reach various values. For nanosecond glow discharges, the increase of temperature and pressure is negligible [6]. For non-thermal sparks, the temperature increases by a few thousand Kelvin and the pressure by a few bars [7]. For thermal sparks, the temperature increases by a few ten thousand Kelvin and the pressure by up to a few hundred bars [8]. The formation of the discharge kernel occurs on time scales shorter than the expansion timescale $\tau_{\text{rarefaction}}$ and is thus isochoric [9,10]. During the post-discharge phase ($t = 0$ s to $t = \tau_{\text{interpulse}}$), there is no longer an electric field in the discharge column (nor a self-induced magnetic field). The plasma cools down and the species created during the discharge phase recombine. The post-discharge can be divided into the three stages illustrated in Figure 1b and Figure 2:

1. $0 < t < \tau_{\text{rarefaction}}$: a blast wave propagates into the surrounding gas (Figure 2b). The leading shock detaches from the plasma kernel and locally increases the temperature of the ambient gas, which may produce radicals. During this phase, due to the concentric rarefaction wave, the temperature of the kernel decreases from $T_k$ to $T_{\text{rarefaction}}$ and the pressure reaches ambient conditions (~1 atm). Note that $\tau_{\text{rarefaction}}$ is often also called the hydrodynamic timescale in the literature.
2. $\tau_{\text{rarefaction}} < t < \tau_{\text{vortex}}$: recirculation cells may form and expel the hot gas kernel radially (Figures 2c, 2d$_1$, 2e$_1$). Vortices form near the electrode (Figure 2c) and move towards the center (Figure 2d$_1$) due to self-induced velocity [11]. Ultimately, a symmetric torus is formed if the top and bottom vortices are of equal strength (Figure 2e$_1$). Figure 3, taken from [12], illustrates the phenomenon. This flow regime is responsible for the final decrease in temperature from $T_{\text{rarefaction}}$ to $T_{\text{end}}$ (see Figure 1b). If the vortices are of unequal strength, an axial jet may form instead of a radial jet. If the created vorticity is too low, the vortices dissipate, and no torus is generated.
3. $\tau_{\text{rarefaction}} < t < \tau_{\text{diff}}$: if no recirculation cell appears (or in parallel to this), the hot kernel cools by thermal diffusion (Figures 2d$_2$, 2e$_2$). Its temperature decreases from $T_{\text{rarefaction}}$ to $T_{\text{end}}$ (see Figure 1b).

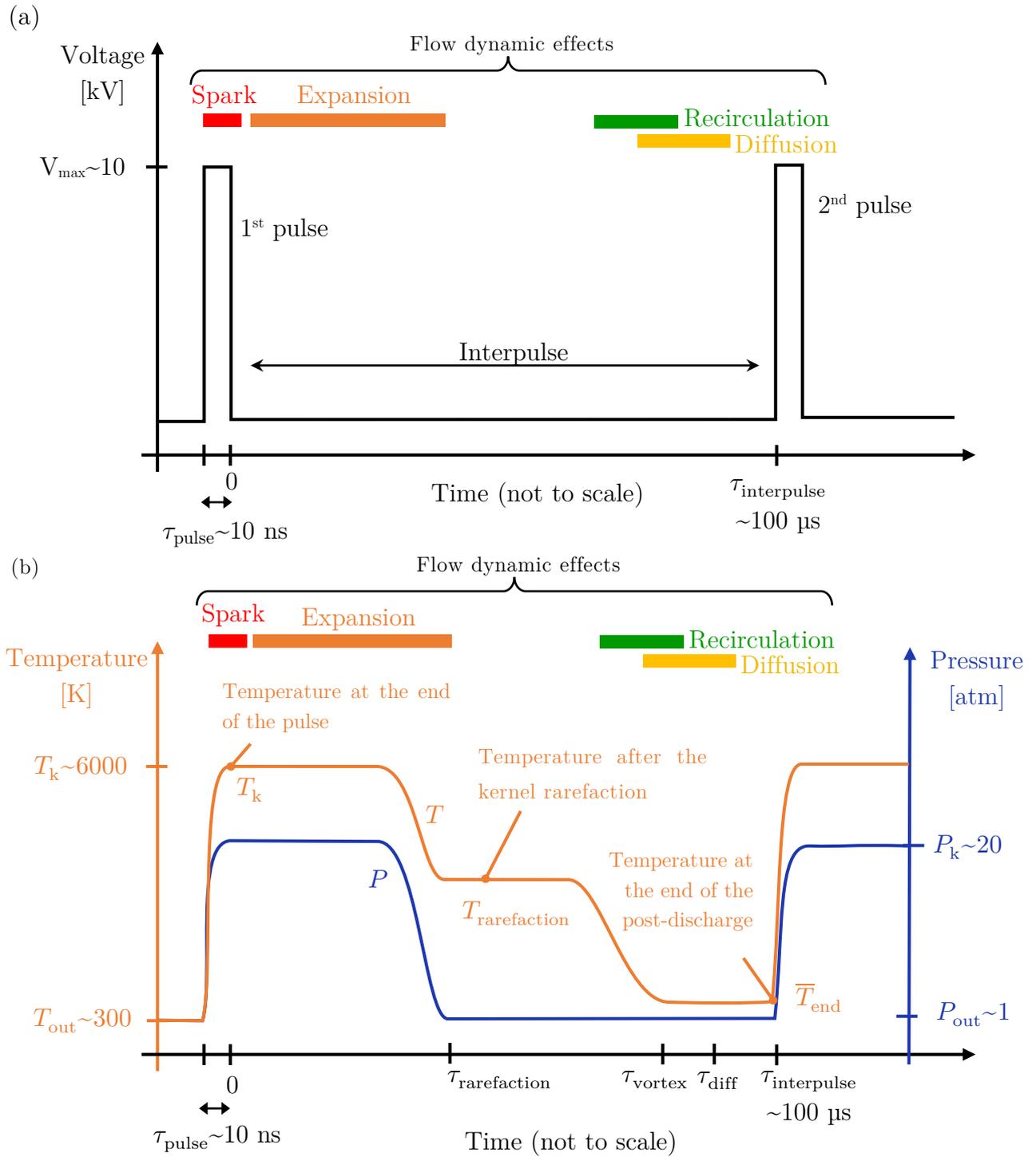

Figure 1. Applied voltage (a) and illustrative mean temperature/pressure profiles in the interelectrode region (b).

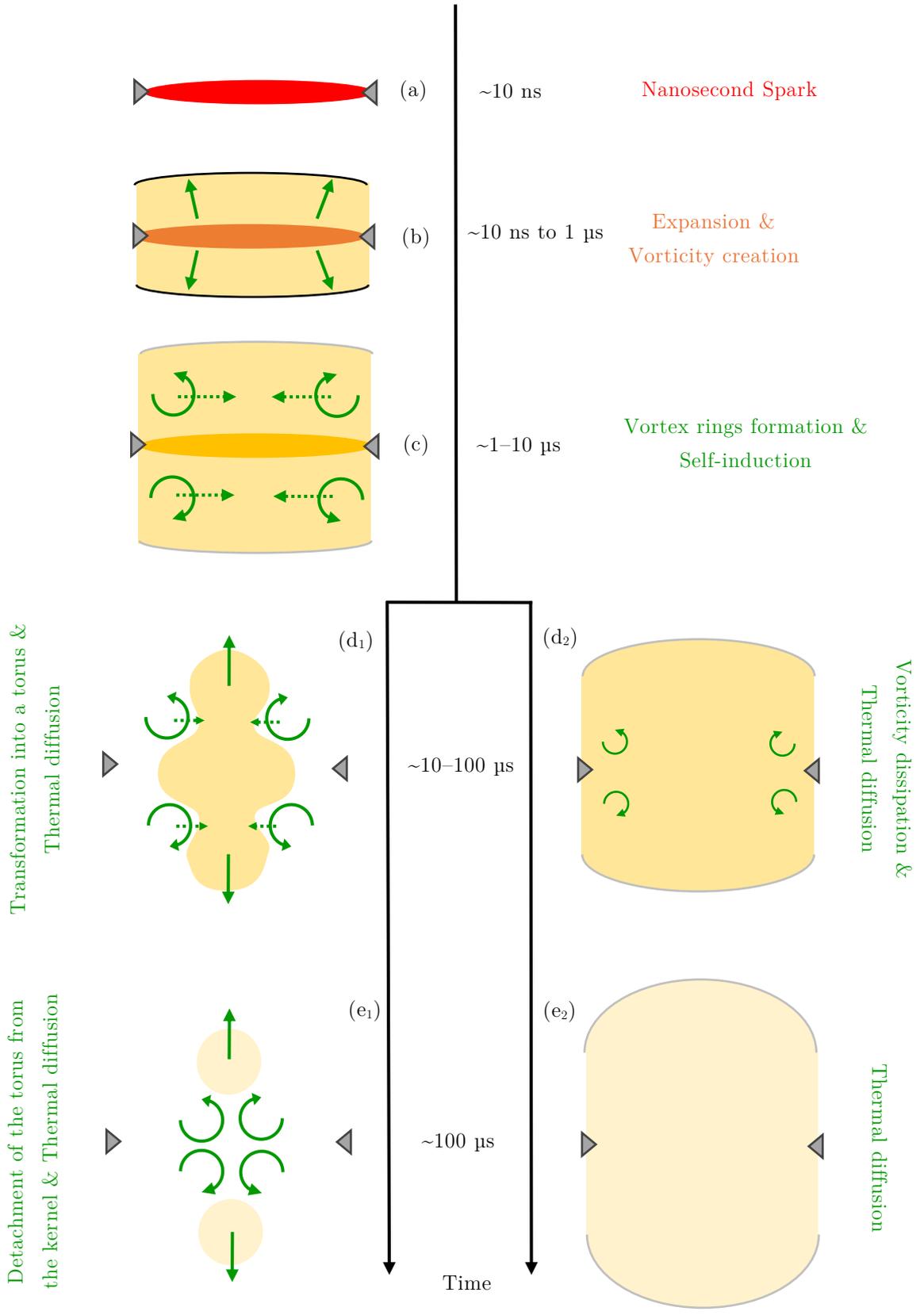

Figure 2. Evolution of the plasma kernel during the post-discharge. Steps (d₁) and (e₁) occur if the post-discharge is in the recirculating regime. Steps (d₂) and (e₂) occur if the discharge is in the diffusive regime (see text).

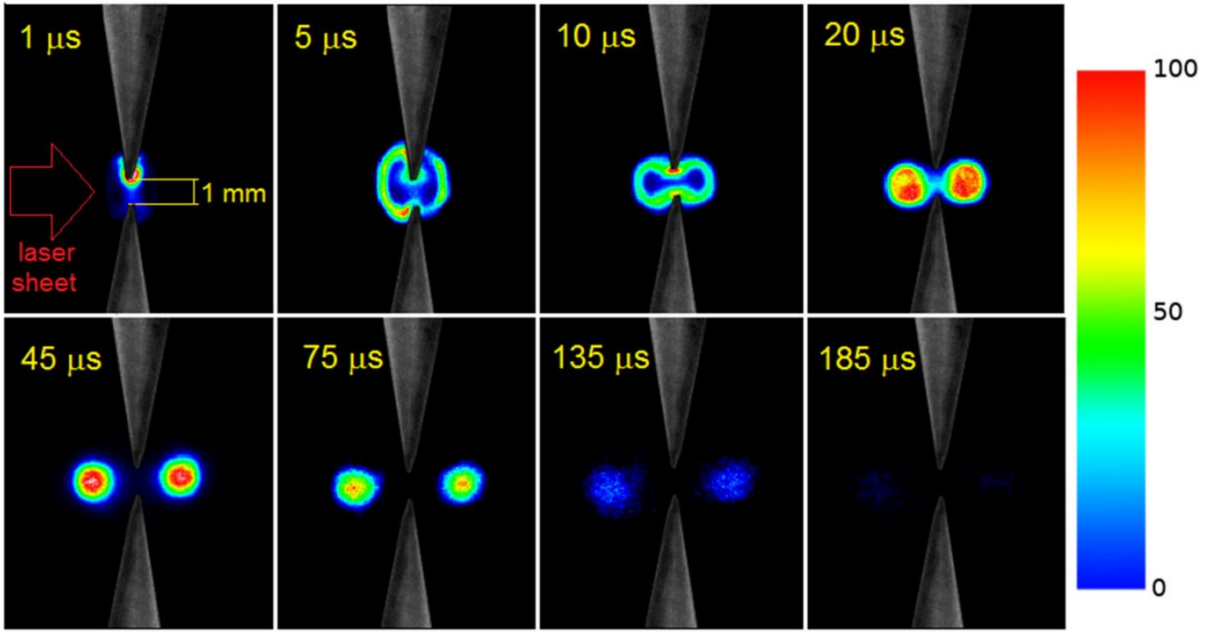

**Figure 3.** Phase-locked OH-PLIF images recorded at different moments in the afterglow of a single nanosecond spark initiated in ambient humid air. The discharge was initiated between pin electrodes by a 10 ns pulse of 30 kV amplitude. Interelectrode gap: 1 mm. Energy deposited in the spark: 1.5 mJ. Camera gate width: 100 ns. The images are taken from [12].

However, there is no consensus in the literature regarding the origin of this recirculation. Some studies claim that the recirculation is caused by a quasi-instantaneous jump of vorticity induced by the non-uniform leading shock [13]. Others consider that this recirculation is mainly linked to the barocline torque, also induced by the shock [14–16]. Yet other publications attribute the recirculation to viscous effects in the boundary layers [17]. Thus, there is a need to clarify the dominant source of vorticity in plasma spark discharges. This will be useful for better understanding the physics and optimizing the design of future industrial reactors for cooling or mixing purposes (by changing the frequency of the pulses, their energy, or the interelectrode gap length). This study has three main objectives: (1) to identify the cooling mechanisms occurring during the interpulse, (2) to determine the primary vorticity sources in the post-discharge, (3) to derive a non-dimensional number $\Pi^*$ to predict the flow regime.

In Section II, we present the numerical model with which we perform CFD (Computational Fluid Dynamics) simulations of the post-discharge for tens of configurations (varying the gap length and the initial kernel temperature). The working gas is air. The simulations are initialized with the post-pulse conditions (since we do not model the pulse phase).

In Section III, we examine the kernel cooling mechanisms (expansion, recirculation, diffusion) predicted in the various CFD simulations. We identify seven different cooling regimes depending on the gap length, initial heating, and repetitive discharge frequency.

In Section IV, we review the sources of vorticity. We use the 2D velocity, pressure, and density fields provided by the CFD simulations to estimate and compare the magnitude of these sources along the central vortex trajectory. We show that the dominant mechanism is the quasi-instantaneous jump of vorticity induced by the non-uniform leading shock. We also demonstrate that the quasi-instantaneous jump of vorticity depends on the ratio of the initial kernel temperature to the ambient temperature.

In Section V, we build a new, physics-based, dimensionless number $\Pi^*$ to characterize the flow that emerges during the interpulse. Through a literature review, we develop a criterion based on $\Pi^*$ to predict the flow regime (recirculating or not).

## II. Description of the numerical model

The fluid dynamic resulting from the discharge is simulated using the commercial Ansys® Fluent® code, release 22.1.0 [18]. We solve the compressible 2D non-reactive Navier-Stokes equations (Eqns. (1)–(7)), in the computational domain represented in Figure 4. Only a quarter of the domain is simulated because the configuration is axisymmetric and has a plane of symmetry at $z = 0$.

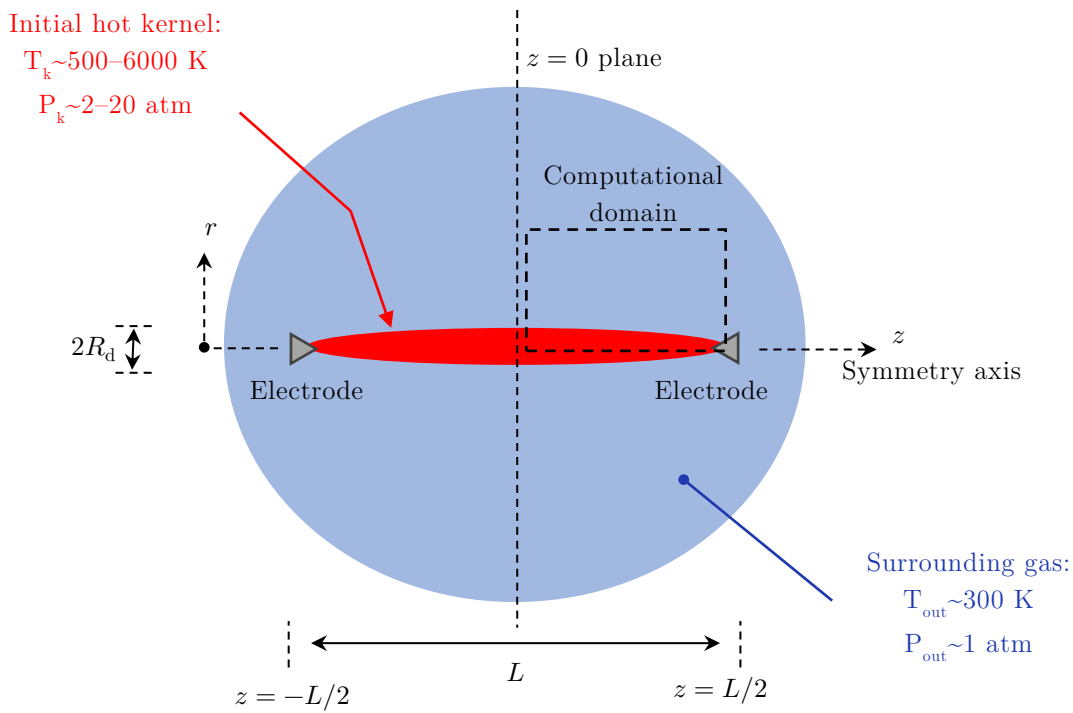

Figure 4. Computational domain. $R_d$ and L represent the radius and the length of the plasma kernel, respectively.

$$\frac{\partial \rho}{\partial t} + \vec{\nabla} \cdot (\rho \vec{u}) = 0 \tag{1}$$

$$\frac{\partial (\rho \vec{u})}{\partial t} + \vec{\nabla} \cdot (\rho \vec{u} \vec{u}) = -\vec{\nabla} P + \vec{\nabla} \cdot \underline{\underline{\tau}} \tag{2}$$

$$\frac{\partial (\rho E)}{\partial t} + \vec{\nabla} \cdot \left( \vec{u}(\rho E + P) \right) = \vec{\nabla} \cdot (\lambda \vec{\nabla} T + \underline{\underline{\tau}} \cdot \vec{u}) \tag{3}$$

$$P = \rho \frac{R}{\mathcal{M}} T \tag{4}$$

$$E = h - \frac{P}{\rho} + \frac{u^2}{2} \tag{5}$$

$$\underline{\underline{\tau}} = \mu \left( (\vec{\nabla} \vec{u} + \vec{\nabla} \vec{u}^T) - \frac{2}{3} (\vec{\nabla} \cdot \vec{u}) \underline{\underline{I}} \right) \tag{6}$$

$$h = h^\circ + \int_{T_{\text{ref}}}^{T} c_p(T') dT' \tag{7}$$

In the previous equations, $\rho$ is the density of the gas, $\vec{u}$ the velocity, $P$ the pressure, $T$ the temperature, $E$ the total specific energy, $\underline{\underline{\tau}}$ the stress tensor, $\underline{\underline{I}}$ the identity tensor, $h$ the specific enthalpy, $h^\circ$ the standard enthalpy of formation at the reference temperature $T_{\text{ref}}$, $\mathcal{M}$ the molar mass, $R$ the ideal gas constant. $c_p$, $\lambda$ and $\mu$ are the specific heat capacity, the thermal conductivity, and the dynamic viscosity, respectively.

After completing a grid convergence study, we selected an unstructured mesh with an average cell size of 2 µm with clustering near the pin-shaped electrodes. The electrode walls are assumed to be adiabatic with a no-slip boundary condition. A finite-volume method is employed to solve Eqns. (1)–(7). Convective flux calculations are based on the Roe solver [19,20], and spatial reconstruction at the cell interfaces is achieved with a third-order MUSCL scheme [21] that blends a central differencing scheme and a second-order upwind scheme with slope limiters [22]. Diffusive flux calculations are performed with a second-order central differencing scheme. We use the Green-Gauss node-based method [23,24] to evaluate the gradients for the two differencing schemes. Finally, time integration is performed using a first-order implicit method [25]. The simulations are performed up to t = 200 µs, with time steps progressively increasing from 50 ps to 10 ns. Thus, the simulations capture the interpulse flow physics for any repetitive discharge with a frequency above 5 kHz.

Initially, the plasma kernel is assumed to be uniform at a temperature $T_k$ and pressure $P_k$. Its shape is an ellipse with a semi-major axis of $L/2 + 10^{-4}$ and a semi-minor axis of $R_d = 100$ µm, where $R_d$ is the discharge radius (see Figure 4). This elliptical shape is inspired by the streamer simulations done in [26]. We define the parameter $r_\mathrm{d}(z)$ (see Eqn. (8)), such that the kernel is the region where $z \leq L/2 + 10^{-4}$ and $r \leq r_\mathrm{d}(z)$.

$$r_\mathrm{d}(z) = R_\mathrm{d} \sqrt{1 - \left(\frac{z}{L/2 + 10^{-4}}\right)^2} \qquad (8)$$

The initial temperature field is thus given by Eqn. (9). To ensure a continuous transition between the hot kernel at $T_\mathrm{k}$ and the cold ambient gas at $T_\mathrm{out}$, we use the following approximation of a step function:

$$T(r,z,t=0) = T_\mathrm{out} + \frac{(T_\mathrm{k} - T_\mathrm{out})}{\left(1 + \left(\frac{r}{r_\mathrm{d}(z)}\right)^{40}\right)\left(1 + \left(\frac{z}{L/2 + 10^{-4}}\right)^{40}\right)} \qquad (9)$$

The power 40 in the denominator controls the steepness of the transition at the edge of the kernel. We chose this value to model a steep transition.

The pulse heating is assumed to be isochoric because its duration ($\tau_\mathrm{pulse}$) is usually much shorter than the expansion timescale ($\tau_\mathrm{rarefaction}$). In addition, we neglect the change of molar mass because the gas does not significantly dissociate or ionize in the temperature range considered in this article ($T_\mathrm{k}$ = 500–6000 K). With these assumptions, the initial pressure and temperature profiles are related as follows:

$$P(r,z,t=0) = T(r,z,t=0) \times \frac{P_\mathrm{out}}{T_\mathrm{out}} \qquad (10)$$

The air heated by the discharge is assumed to be in local thermodynamic equilibrium (LTE) throughout the post-discharge. The thermodynamic and transport properties ($c_p$, $\lambda$, $\mu$) are taken from [27].

The various assumptions made for the simulations (kernel shape, kernel radius, electrode shape, steepness of the transition at the edge of the kernel, LTE) are further discussed in Appendix A. We make new CFD simulations, and we show that these assumptions do not significantly influence the cooling of the kernel and do not alter the conclusions made in this article, except if the electrodes are planar.

### III. Description of the interpulse flow and identification of cooling regimes

In this Section, we perform simulations of the post-discharge varying the gap length $L$ from 1 to 5 mm and the initial temperature of the kernel $T_\mathrm{k}$ from 500 to 6000 K, typical of glow discharges or non-thermal sparks. First, we provide a detailed description of the flow from four characteristic simulations corresponding to parameters $(L, T_\mathrm{k})$ at the edge of the domain explored. Second, we extend the analysis to tens of simulations and show that the cooling of the discharges can be classified into seven regimes (labeled A–G).

## III.A. Description of different characteristic flows

In this Subsection, we focus on the following four configurations:

1. $L = 1$ mm, $T_k = 5500$ K
2. $L = 5$ mm, $T_k = 5500$ K
3. $L = 1$ mm, $T_k = 500$ K
4. $L = 5$ mm, $T_k = 500$ K

The four cases are representative of glow and spark nanosecond discharges across typical gap lengths used in experiments (see Table 1 of [6]). For each simulation, we calculate the mean temperature $\bar{T}$ along the interelectrode axis (see Eqn. (11)). The results are compared in Figure 5. In addition, the temperature, velocity, and azimuthal vorticity[b] fields at different timesteps are shown in Figures 6–9, respectively in the left, middle, and right columns of each Figure. In the following paragraphs, we analyze the cooling phases of each case, using Figure 5. In parallel, we study the flow behind these different phases with more details in Figures 6 to 9.

$$\bar{T}(t) \equiv \frac{1}{L/2} \int_{z=0}^{z=L/2} T(r=0, z, t) dz \qquad (11)$$

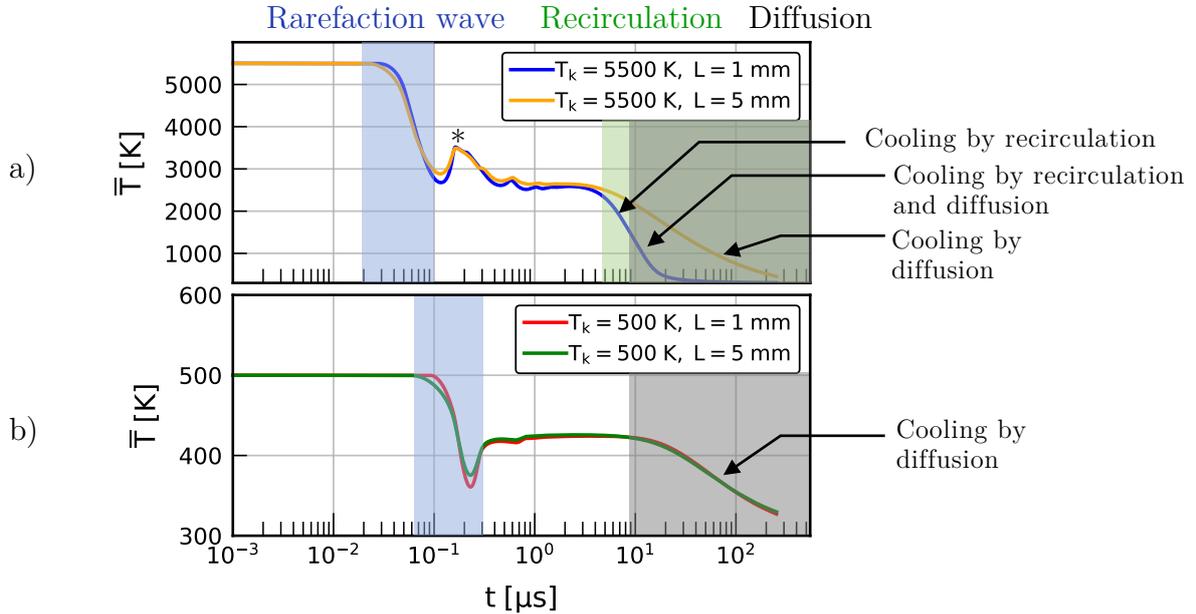

Figure 5. Time history of the average temperature computed along the pin-to-pin axis in the three cooling phases (rarefaction, recirculation, diffusion). (a) Cases 1 and 2. (b) Cases 3 and 4. The asterisk ($*$) marks the temperature peak due to the secondary shock (see text).

---

[b] $\omega \equiv (\vec{\nabla} \times \vec{u}) \cdot \vec{e_\theta}$ with $\vec{e_\theta}$ the azimuthal unit vector.

In Case 1 ($L = 1$ mm, $T_k = 5500$ K), the blue curve of Figure 5a shows that the temperature of the plasma kernel decreases abruptly between 0.02 and 0.1 µs due to the rarefaction wave, reaching approximately half of its initial value at $\tau_\text{rarefaction} = 0.1$ µs. A secondary shock (see [28] for theoretical details) increases the temperature by more than 500 K between 0.1 and 0.2 µs (this is marked by an asterisk in Figure 5a). The temperature relaxes to a plateau 0.2 µs later. At 0.5 µs, Figure 6 shows the expanding leading shock, characterized by a non-uniform curvature. At 10 µs, the temperature field reveals a prominent clockwise recirculation vortex centered on point B, bringing cold gas into the interelectrode region and expelling the hot plasma kernel axially along the symmetry plane. The vorticity field indicates that this clockwise recirculation pattern (blue) dominates other counterclockwise recirculation vortices (red) and originates from a region centered on point A in the vorticity snapshot at 500 ns. This region is located near the electrode, and the clockwise vortices appear after the leading shock front has passed. The vortex ring has a self-induced velocity [11] directed towards the center of the gap. Subsequently, the vortex moves toward point C, transporting the gases along the interelectrode axis, trapping them close to the interelectrode plane, and forcing them to evacuate radially [29]. The vortex is responsible for the cooling between 5 and 20 µs shown in Figure 5a (blue curve). This process is summarized in Figure 10.

In Case 2, the interelectrode gap length is 5 mm, five times longer than in Case 1. The orange curve of Figure 5a shows that the post-rarefaction temperature decrease is slower and occurs between 5 and 300 µs. In Figure 7, we still observe the propagation of a leading shock wave, the generation of clockwise vorticity near the electrode (point A), and the displacement of the vortex towards the center along the path A-B-C. However, at $t = 80$ µs, the center of the vortex has only traveled about a quarter of the interelectrode gap distance (point C) and has not reached the middle plane. Although the same amount of vorticity was created initially at point A, the vortex needs more time to transport fresh gas to the symmetry plane in Case 2 than in Case 1. This situation corresponds to stages (a)–(c) in Figure 10. During this longer travel, three phenomena affect the kernel cooling:

- Thermal diffusion becomes significant at timescales around tens of microseconds and adds its cooling effect to the convective cooling mechanism described above.
- The vortex dissipates energy via viscous effects, decelerates, and may even vanish before reaching the symmetry plane. In such cases, only diffusion remains to slowly cool the plasma kernel. This corresponds to the diffusive regime shown in Figure 2.
- Breakdown of the next pulse: if the frequency is high enough, the cooling processes may be interrupted by a new pulse, which heats back the interelectrode region.

Cases 3 and 4 correspond to the same gap lengths as in Cases 1 and 2, except that now the initial temperature of the plasma kernel is set to 500 K instead of 5500 K. Figures 8 and 9 show that the amount of vorticity is much smaller and no recirculation cell appears. The temperature decreases because of diffusion only (green and red curves in Figure 5b) between 5 and 300 µs (on the same time scale as in Case 2). This corresponds also to the diffusive regime shown in Figure 2 and discussed in Dumitrache *et al.* [15].

As far as we know, there are no quantitative temperature measurements available to validate the predictions of the CFD model shown in Figure 5 and Figures 6–9 during the full interpulse. Indeed, these measurements are complex and usually stop before the gas has recirculated [9]. However, we can make a preliminary comparison between the temperature snapshots of Figure 6 and the OH-PLIF images recorded by Stepanyan *et al.* [12] and shown in Figure 3. We observe a similar collapse into a toroidal shape of the temperature field and the fluorescence signal. This similarity gives us confidence that our numerical model represents the post-discharge behavior following a nanosecond pulse.

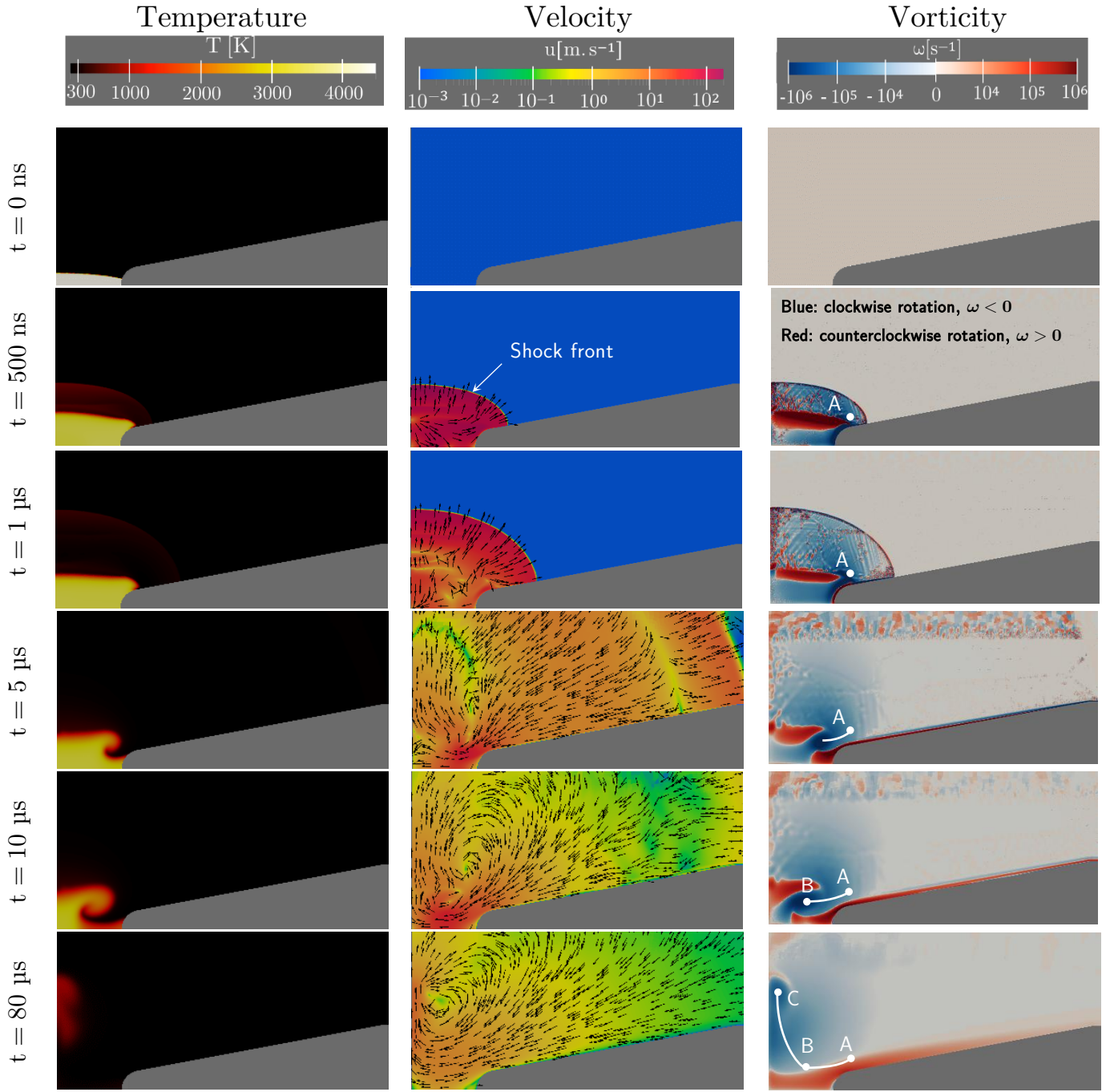

**Figure 6.** Computed temperature (left), velocity (center), and vorticity (right) fields for a discharge with a 1 mm gap length and a 5500 K initial plasma kernel (case 1). In the right-hand column images, blue regions indicate clockwise (negative) vorticity, and red regions indicate counterclockwise (positive) vorticity. The trajectory of the center of the main clockwise vortex (taken as the point of maximum clockwise vorticity) is represented by the white path A-B-C.

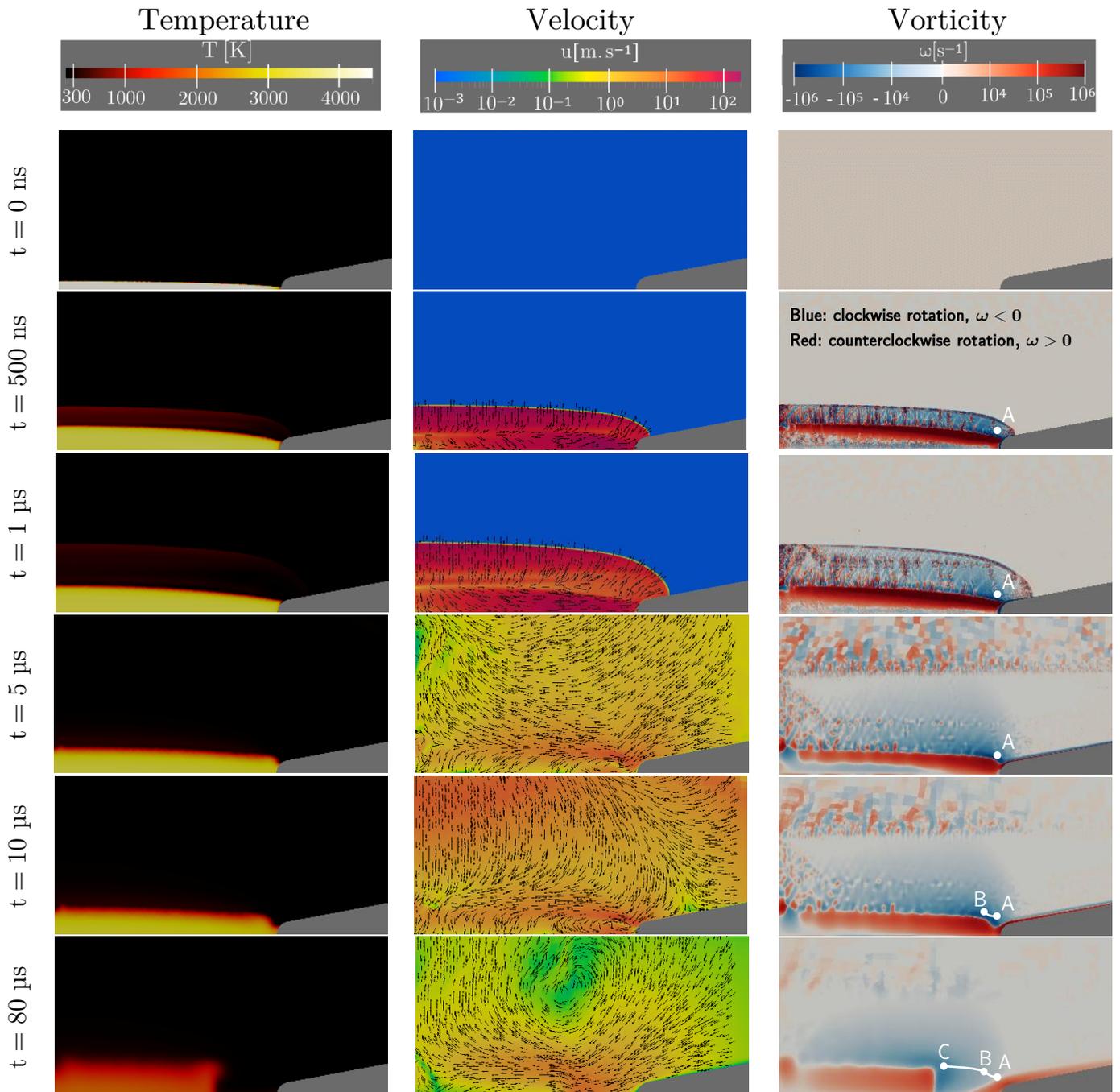

**Figure 7.** Computed temperature (left), velocity (center), and vorticity (right) fields for a discharge with a 5 mm gap length and a 5500 K initial plasma kernel (case 2). In the right-hand column images, blue regions indicate clockwise (negative) vorticity, and red regions indicate counterclockwise (positive) vorticity. The trajectory of the center of the main clockwise vortex (taken as the point of maximum clockwise vorticity) is represented by the white path **A-B-C**.

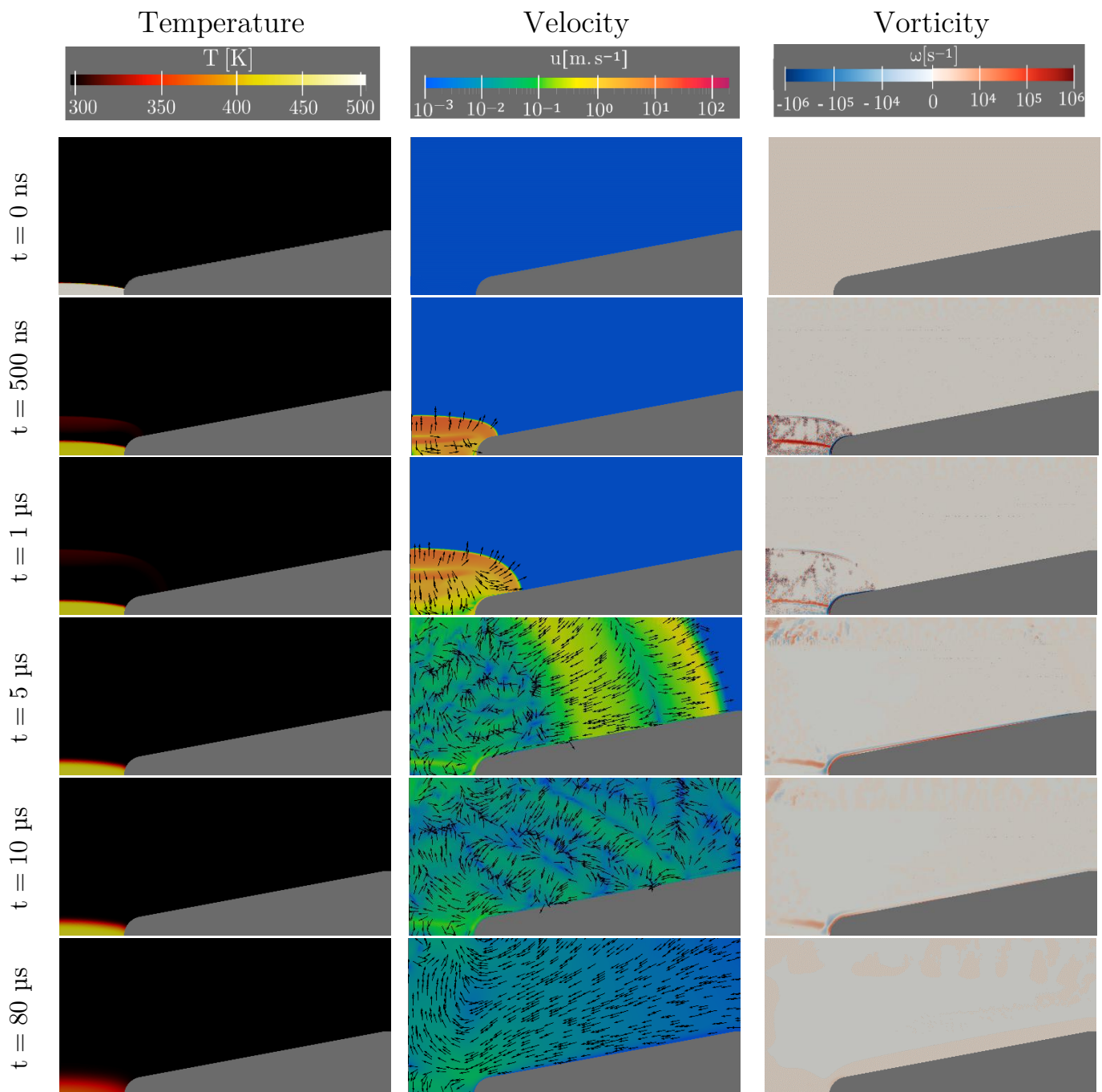

**Figure 8. Computed temperature (left), velocity (center), and vorticity (right) fields for a discharge with a 1 mm gap length and a 500 K initial plasma kernel (case 3).**

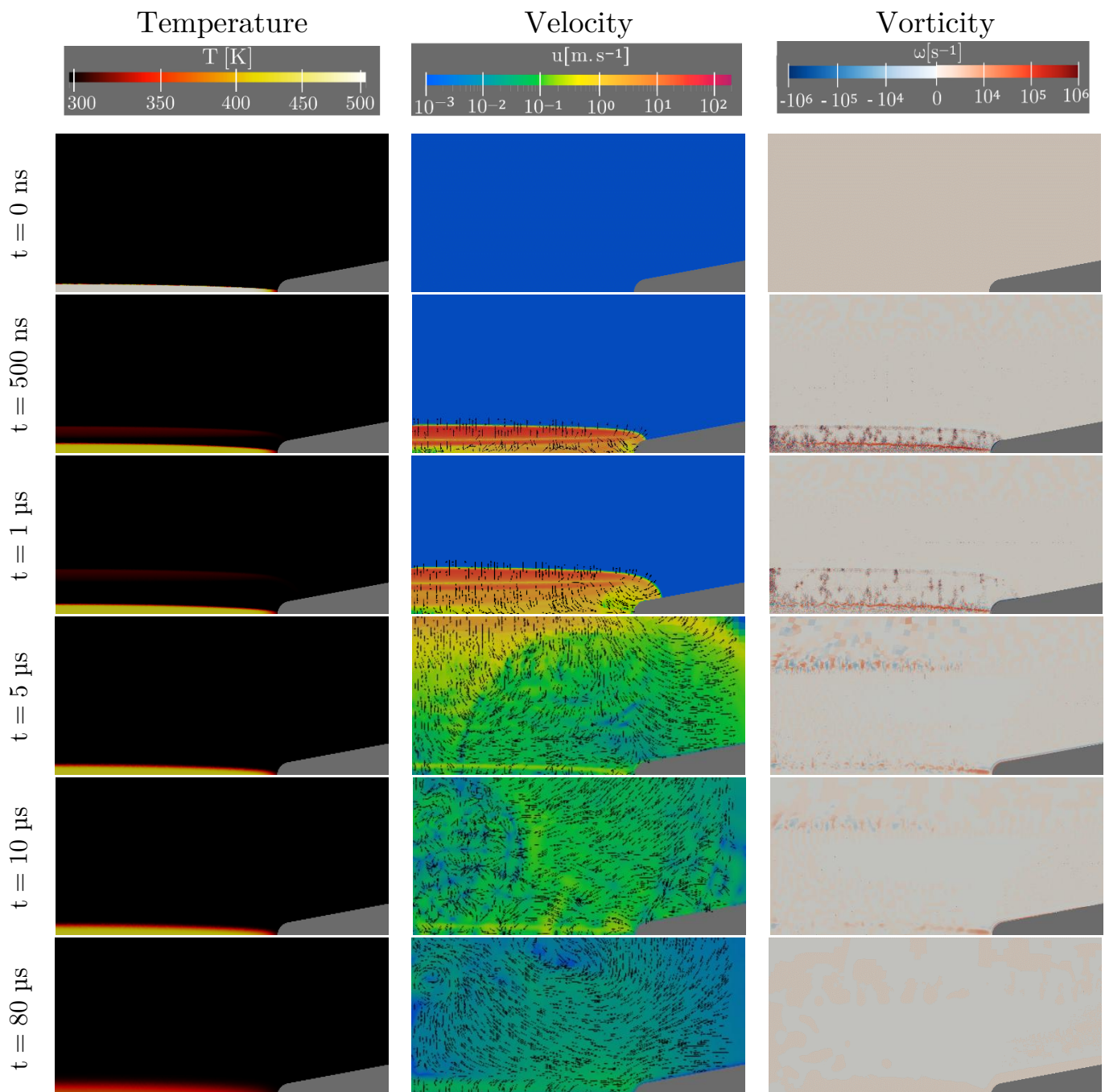

**Figure 9.** Computed temperature (left), velocity (center), and vorticity (right) fields for a discharge with a 5 mm gap length and a 500 K initial plasma kernel (case 4).

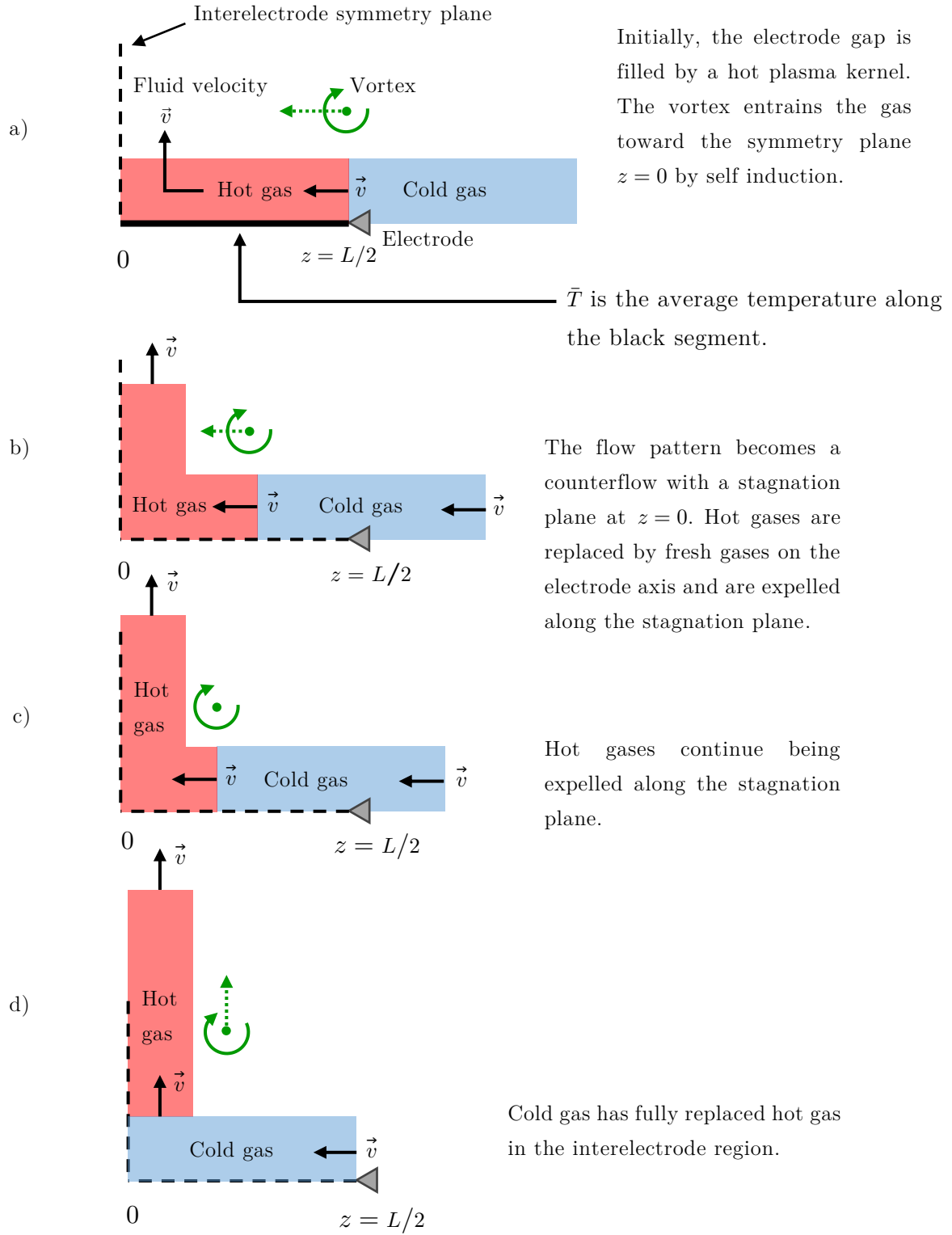

**Figure 10.** Schematic of the replacement of hot gas by cold gas in the discharge interelectrode region as a result of a recirculation cell. The green dashed arrow represents the displacement vector of the vortex center.

## III.B. Identification of the main cooling regimes

In this Subsection, we expand our analysis to all the CFD simulations performed with gap lengths $L$ ranging from 1 mm to 5 mm and initial kernel temperatures $T_k$ ranging from 500 K to 6000 K. Additionally, we consider different pulse frequencies $f$ spanning the range 5–100 kHz. For each set of discharge parameters $(T_k, L, f)$, we calculate the mean temperature along the interelectrode axis just before the next pulse $\bar{T}_{\text{end}}$ (see Eqn. (12) and Figure 1b).

$$\bar{T}_{\text{end}} \equiv \bar{T}(t = 1/f) \tag{12}$$

This average temperature $\bar{T}_{\text{end}}$ serves as a simple scalar indicative of the recirculation regime. Before examining the results, several cases can be identified in relation to what we previously described in Figure 5 and Figure 10:

- $\bar{T}_{\text{end}} \approx T_k$ means that neither expansion, recirculation, nor diffusion can significantly cool the kernel before the next pulse. This case does not occur in our simulations. It would only occur if the frequency were high enough ($f^{-1} > \tau_{\text{rarefaction}}$).
- $\bar{T}_{\text{end}} \approx T_k/2$ means that recirculation and diffusion have no effect on the cooling of the kernel before the next pulse. The temperature decreases only through the expansion process (Figure 5a).
- $T_{\text{out}} \lesssim \bar{T}_{\text{end}} \lesssim T_k/2$ means that the cooling is incomplete. Either the recirculation cell did not have enough time to expel all the hot gas (as schematized in Figure 10c), and/or diffusion was still in progress when the next pulse started.
- $\bar{T}_{\text{end}} \approx T_{\text{out}} = 300$ K indicates that complete cooling is obtained thanks to expansion, recirculation (see Figure 10d) and/or diffusion. The kernel reaches the temperature of the surrounding gas.

We now examine the values of $\bar{T}_{\text{end}}$ as a function of $T_k$, $L$, and $f$.

First, Figure 11a shows $\bar{T}_{\text{end}}$ for different values of the initial kernel temperature $T_k$ and gap length $L$, at a given frequency $f = 40$ kHz. Four regimes can be identified. In regime A, corresponding to the top left-hand corner, we have $\bar{T}_{\text{end}} \approx T_{\text{out}}$. This is the regime of case 1 ($T_k = 5500$ K, $L = 1$ mm), for which the recirculation is sufficiently quick and strong to entrain cold gases along the pin-to-pin axis. If we increase the gap length, we reach regime B, where the vortices are not fast enough to travel over the entire gap during the interpulse duration $1/f$. The cooling is only partial, mainly due to expansion but also partly to diffusion. If the gap length is increased, the final temperature increases because the fraction of gap length crossed by the vortices decreases. Interestingly, the separation between regime A and regime B is not vertical: increasing the kernel temperature increases the strength of the vortices, which move faster and can cross a longer portion of the gap. The link between the kernel temperature and the amount of vorticity generated is discussed in Section IV.C. If the initial kernel temperature is relatively low (below 1000 K), the kernel completely cools ($\bar{T}_{\text{end}} \approx T_{\text{out}}$) regardless of the gap length: this is characteristic of the absence of recirculation. Cooling occurs only by expansion and radial diffusion. We refer to this regime as D. In regime C, the final temperature does not fully equilibrate with the ambient temperature before the following pulse. In addition, $\bar{T}_{\text{end}}$ does not depend on the gap length, hence the main cooling mechanisms of regime C are also expansion and diffusion.

Second, Figure 11b shows $\bar{T}_{\text{end}}$ for different values of the frequency and gap length at a given initial temperature $T_{\text{k}} = 3300$ K. At high frequencies (100 kHz), the interpulse is very short, and only expansion has enough time to occur, partially cooling the kernel. The final temperature is approximately half of the initial one, regardless of the gap length. We call this regime E. In the bottom left-hand corner, we fall back into regime A. The interpulse is sufficiently long for the vortices to travel the entire gap length. The kernel can fully cool even at long gaps if the frequency is sufficiently low. The rest of the domain can be split into approximately three zones. If the frequency is too high for the diffusion to occur, we are in regime B. Otherwise, the three phenomena (expansion, recirculation, and diffusion) are all significant and contribute to kernel cooling. We distinguish regime F, where the cooling is partial, and regime G, where the cooling is complete.

Third, Figure 11c shows $\bar{T}_{\text{end}}$ for different values of the frequency and initial temperature at a given gap length $L = 2$ mm. In this parametric space, regime identification is more complex. We suggest an approximate mapping of the regimes A–G. We remark that at a fixed frequency of 20 kHz, a post-discharge initially at 6000 K cools completely and reaches 300 K, unlike a post-discharge initially at 2000 K. This difference in cooling efficiency is attributed to a difference in the strength of the recirculation cell. The blast wave generates more vorticity in the hotter scenario, leading to faster flow recirculation. Consequently, the hotter scenario allows enough time for the cold gases to be brought along the interelectrode axis. We provide a physical explanation of this phenomenon in the following Section.

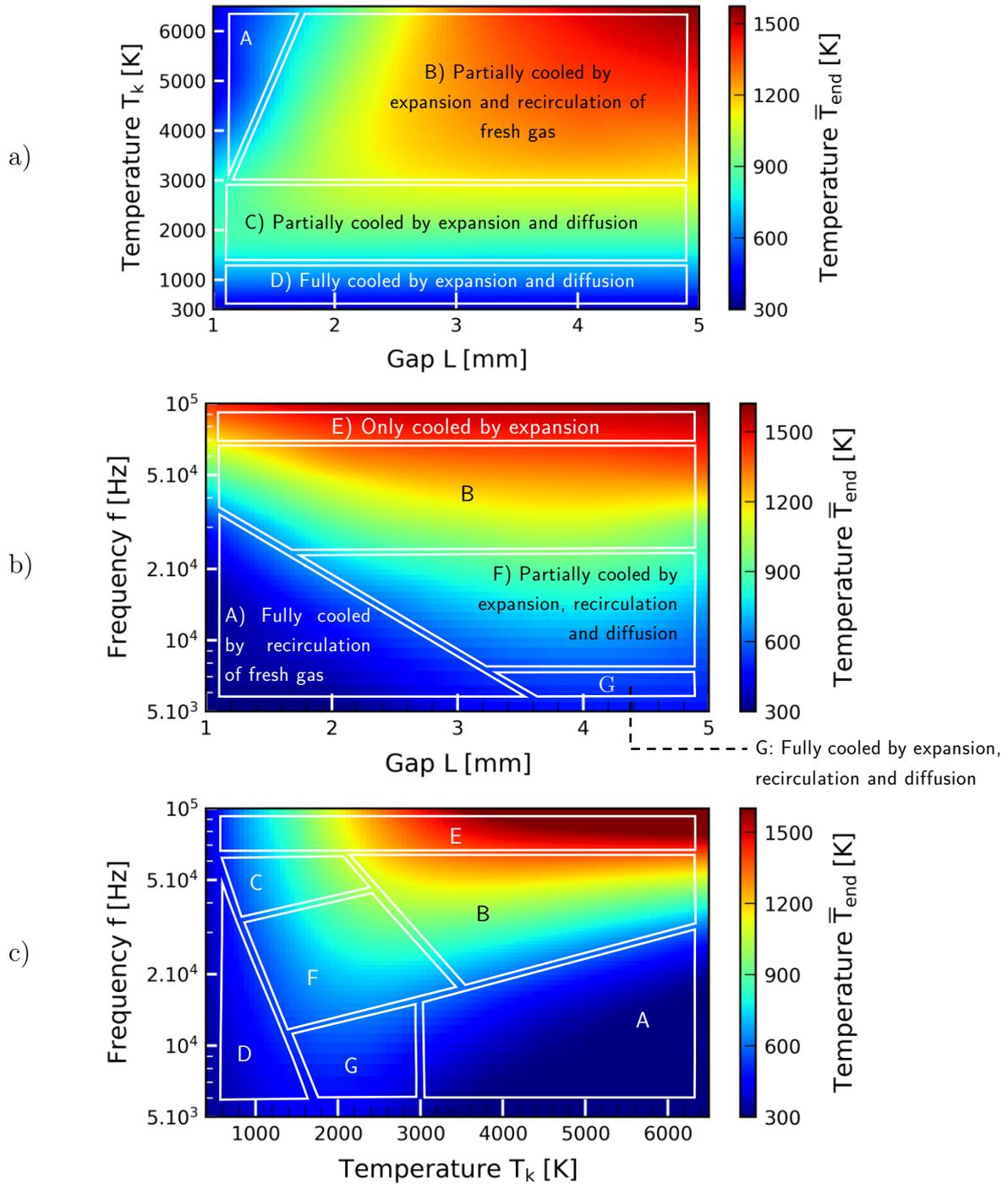

Figure 11. Mean temperature $\bar{T}_{\mathrm{end}}$ along the pin-to-pin axis at the end of the interpulse for different sets of discharge parameters $(L, f, T_{\mathrm{k}})$. Cooling regimes A to G characterize the dominant cooling mechanisms. (a) Influence of $L$ and $T_{\mathrm{k}}$ on $\bar{T}_{\mathrm{end}}$; $f$ is fixed to 40 kHz. (b) Influence of $L$ and $f$ on $\bar{T}_{\mathrm{end}}$; $T_{\mathrm{k}}$ is fixed at 3300 K. (c) Influence of $T_{\mathrm{k}}$ and $f$ on $\bar{T}_{\mathrm{end}}$; $L$ is fixed to 2 mm.

# IV. Vorticity generated by a nanosecond pulse

In this Section, we quantitatively analyze the vorticity sources that govern the flow dynamic regimes (recirculating or not) and influence the various cooling regimes discussed in Section III. In Subsection IV.A, we provide a theoretical review of the physical processes driving the different sources of vorticity. In Subsection IV.B, we compute and compare the sources of vorticity along a particle trajectory that follows the center of the vortex responsible for radially expelling the hot gas and cooling the kernel. In Subsection IV.C, we show that the amount of vorticity generated depends on the ratio of the initial kernel temperature to the ambient temperature.

## IV.A. Theory of vorticity generation

The vorticity change is given by the vorticity balance equation [30,31]. Since our configuration is axisymmetric, we only focus on the azimuthal component of vorticity $\omega \equiv (\vec{\nabla} \times \vec{u}) \cdot \vec{e_\theta}$. In a Eulerian description, the balance equation reads:

$$\frac{\partial \omega}{\partial t} = \underbrace{S_\text{advection}}_{\substack{\text{convective} \\ \text{flux of} \\ \text{vorticity}}} + \underbrace{S_\text{barocline} + S_\text{viscous} + S_\text{stretching} + S_\text{expansion}}_{\text{source terms for the continuous regions of the flow}} + \underbrace{S_\text{shock}}_{\substack{\text{source term} \\ \text{caused by} \\ \text{discontinuous} \\ \text{regions of the} \\ \text{flow}}} \quad (13)$$

The first term on the right-hand side of Eqn. (13) corresponds to the convective flux of vorticity from/to other regions of the flow. This term is included in the definition of the Lagrangian/material derivative $d\omega/dt = \partial \omega/\partial t - S_\text{advection}$.

$$S_\text{advection} = -(\vec{u} \cdot \vec{\nabla})\omega \quad (14)$$

The following four terms ($S_\text{barocline}$, $S_\text{viscous}$, $S_\text{stretching}$, $S_\text{expansion}$) correspond to sources and sinks changing the vorticity in continuous flow regions (i.e. everywhere except across a discontinuous shock). They account for changes in vorticity due to pressure and density gradients (the barocline torque, $S_\text{barocline}$), viscous dissipation ($S_\text{viscous}$), vortex stretching ($S_\text{stretching}$) or density changes ($S_\text{expansion}$). These terms read:

$$S_\text{barocline} = \frac{1}{\rho^2}(\vec{\nabla}\rho \times \vec{\nabla}P) \cdot \vec{e_\theta} \quad (15)$$

$$S_\text{viscous} = \vec{\nabla} \times \left(\frac{\vec{\nabla} \cdot \underline{\underline{\tau}}}{\rho}\right) \cdot \vec{e_\theta} \quad (16)$$

$$S_\text{expansion} = -\omega(\vec{\nabla} \cdot \vec{u}) \quad (17)$$

$$S_{\text{stretching}} = \left((\vec{\omega} \cdot \vec{\nabla})\vec{u}\right) \cdot \vec{e_\theta} \qquad (= 0 \text{ in 2D}) \tag{18}$$

These four source terms do not have an instantaneous effect. Each term has its own characteristic time, and it takes time to generate or reduce the amount of vorticity. Therefore, we refer to them as the "non-instantaneous vorticity source terms." Among these, $S_{\text{barocline}}$ was proposed as the most likely cause of the overall recirculation pattern observed in nanosecond post-discharges [14–16]. We refer to this mechanism as the "Barocline Torque" (BT) mechanism.

The last term of Eqn. (13), $S_{\text{shock}}$, corresponds to the vorticity deposited at a specific location by a non-uniform shock that crosses that location. This deposition happens very quickly, about the inverse of the collision frequency, and can be considered instantaneous on the time scales over which the Navier-Stokes equations are valid. This mechanism will be referred to as the "Instantaneous Shock-Induced Vorticity" (ISIV) mechanism. Let us define $t_s$, the time at which a shock front reaches a specific location in the flow. The vorticity jump instantaneously induced by the shock is denoted $\Delta\omega$. $S_{\text{shock}}$ is a scaled Dirac delta function whose expression was derived initially by Truesdell [32] and generalized by Lighthill [33] and then Hayes [30]. In the frame of reference of the shock, the upstream flow velocity has no tangential component because the ambient gas is at rest. Therefore, in our configuration, Hayes's formula can be simplified to:

$$S_{\text{shock}} = \delta(t - t_s) \times \Delta\omega \tag{19}$$

$$\Delta\omega = \left[-\vec{n} \times \left(\vec{\nabla}_t u_{0,n}\right)\left(\frac{\rho_1}{\rho_0} - \frac{\rho_0}{\rho_1} - 2\right)\right] \cdot \vec{e_\theta} \tag{20}$$

where $\delta$ is the Dirac delta function. $\vec{n}$ and $\vec{t}$ are the two unit vectors displayed in Figure 12a. $\vec{n}$ is normal to the shock front and oriented toward the flow. $\vec{t}$ is tangent to the shock front. $\vec{\nabla}_t$ corresponds to the tangential gradient (along the shock front). Index 1 refers to the flow downstream of the shock. Index 0 refers to the flow upstream of the shock. The upstream flow is also denoted by the subscript "out" in the other Sections of this article, as the upstream domain corresponds to the outside domain in the flow configuration we study. $u_{0,n}$ is the normal velocity of the upstream flow relative to the shock front. $\rho_0$ and $\rho_1$ are the flow densities, linked to the shock Mach number by the Rankine-Hugoniot relations [34].

Eqn. (20) states that $S_{\text{shock}}$ will be non-zero only if $\vec{\nabla}_t u_{0,n} \neq 0$, i.e. if the shock strength is non-uniform along the front. As we will demonstrate, NRP discharges induce non-uniform shocks and are very efficient in generating vorticity via $S_{\text{shock}}$ (Eqns. (19)–(20)). The general ISIV mechanism is explained below and illustrated in Figure 12. We explain the vorticity creation near several points (labeled from A to H) by examining how a typical shock front expands over time. We trace the shape of the shock at four different instants using additional dots (labeled $S_0$ to $S_7$) colored according to the local speed of the shock.

a. The discharge creates a hot plasma kernel, as shown in Figure 12a. This kernel is modeled with an elongated ovoid with a small radius of curvature close to the tip of the electrode (typically $2R_d^2/L$ for an elliptical shape) and a large radius of curvature close to the interelectrode symmetry plane (typically $L^2/(4R_d)$ for an elliptical shape). The temperature and the pressure inside the kernel are approximately uniform and significantly higher than the ambient ones. In step (a) (quasi-initial conditions), the kernel is separated from the cold freestream region by an attached leading shock of ovoidal shape and uniform strength. For now, all the dots $S_0$–$S_7$ along the shock are violet (maximal and uniform shock strength).

b. After its creation, the leading shock detaches and propagates into the surrounding medium. Figure 12b represents the instant just before it reaches the points A–D, initially at rest and without vorticity. Between steps (a) and (b), the shock becomes non-uniform. Indeed, its strength decreases faster in $S_0$ and $S_1$ than in $S_6$ and $S_7$ because the expansion of the front is spherical in $S_0$ and $S_1$ and cylindrical in $S_6$ and $S_7$. The strength of a spherical shock is known to decrease more abruptly than for a cylindrical shock because the energy is distributed over a larger surface during the propagation [35]. The situation is intermediate in $S_2$ and $S_3$, but the shock strength is higher in $S_3$ than in $S_2$. Note that the strength is identical in $S_6$ and $S_7$ because both points lie on a quasi-perfectly cylindrical front (the strength decreases in the same proportion).

c. Figure 12c represents the instant shortly after the shock has crossed the points A–D. Points A and B were traversed by the portions of the shock front of strength $S_2$ and $S_3$, respectively, with $S_2 < S_3$. Thus, the post-shock velocities in A and B differ and $\overrightarrow{\nabla} \times \vec{u}$ is no longer zero: vorticity has been induced by the shock. As the change in velocity $\vec{u}$ is sudden, the change in vorticity $\overrightarrow{\nabla} \times \vec{u}$ is also sudden. Unlike A and B, points C and D were traversed by $S_6$ and $S_7$, whose strengths are identical. Hence, the post-shock velocities $\vec{v}_C$ and $\vec{v}_D$ are identical, and no vorticity is produced. The amount of instantaneously induced vorticity will be maximal in regions where the shock front is highly non-uniform, i.e., where the radius of curvature varies the most, near the tip of the electrode.

d. After some time, as illustrated in Figure 12d, the leading shock wave transforms into a uniform spherical shock because the sections of the front with the smallest curvature radius expand slower than the sections with the highest curvature radius. As a result, all sections eventually have the same curvature radius, move at the same speed, and have the same strength. After that moment, the post-shock velocities at all points (E, F, G, and H, for instance) crossed by the shock front have the same magnitude and are all aligned with the center O of the discharge. No additional vorticity is created by the mechanism of step (c).

Note that much confusion is associated with the source term $S_\text{shock}$. First, some studies ignored the ISIV mechanism in their analysis [14,15]. This is a significant issue because ISIV is dominant in many practical applications of nanosecond discharges. Second, certain studies mentioning the ISIV mechanism (and using equations similar to Eqn. (19)–(20)) employed the denomination "barocline torque" [13]. This is a source of confusion because the terms $S_\text{shock}$ and $S_\text{barocline}$ in Eqn. (13) involve two mechanisms with different timescales that should be named differently. The present work aims to remove the confusion and clarify the situation regarding vorticity generation by nanosecond discharges. In summary, shock waves have two main effects on the vorticity of a fluid:

- The first effect is instantaneous, resulting in a discontinuous change of vorticity by the ISIV mechanism, which can be represented by $S_\text{shock}$ given in Eqn. (13).
- The second effect is not instantaneous and leads to a gradual change in vorticity. It takes some time for this effect to impact the flow field. During the passage of the shock, the density, velocity, and pressure undergo sudden changes. At this point, the gas may be in a state where certain source terms ($S_\text{barocline}$, $S_\text{viscous}$, $S_\text{stretching}$, $S_\text{expansion}$, $S_\text{advection}$) are no longer zero. Although these source terms change suddenly, their impact on the vorticity occurs gradually over time.

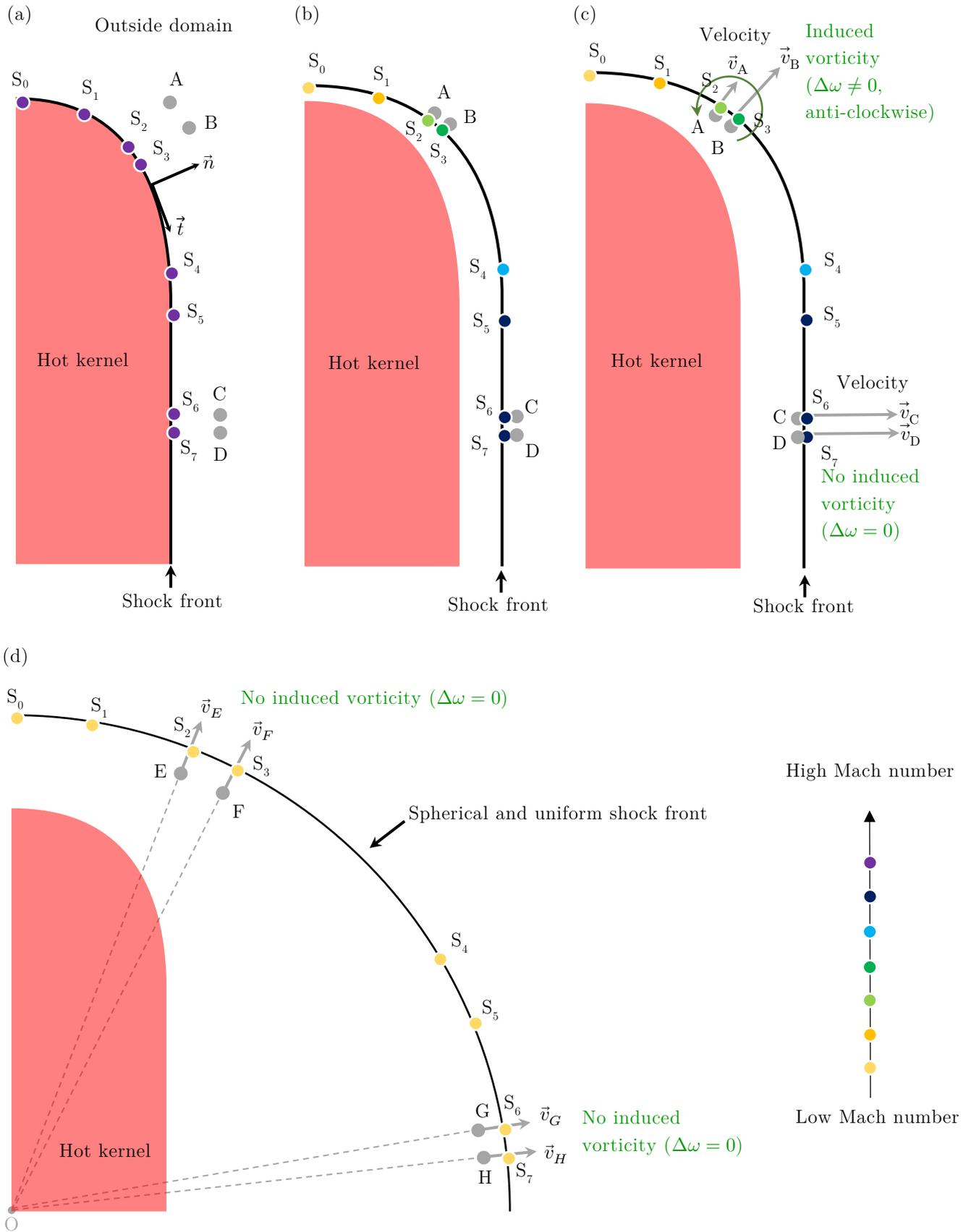

Figure 12. Sketch showing the role of the leading shock in instantaneous vorticity creation. It illustrates the physical phenomena modeled by $S_{\text{shock}}$ in Eqn. (13). See text for details.

## IV.B. Comparison of the different vorticity sources during the post-discharge

We now aim to determine the contribution of the various source terms in Eqn. (13) along the trajectory of the principal vortex that drives the overall recirculation pattern.

The analysis is performed using the CFD simulation corresponding to the case $T_\mathrm{k} = 6000$ K and $L = 3$ mm, which exhibits a clear recirculation cooling pattern. As we will see in Figure 13, the principal vortex (responsible for expelling the hot gas kernel radially and generating the torus) originates from a region near the electrode tip. This point is indicated by the white circle in Figure 13. We place a probe point at that location. This probe point represents an element of fluid convected by the flow. Its integrated trajectory is indicated by the white line in Figure 13. As expected, the probe point is always located inside the core of the vortex. The vorticity of the flow, computed at each instant at the location of the probe, is presented in Figure 14a. We identify three phases:

- Between $t = 0$ and 0.2 µs, the flow is at ambient conditions and is not rotating ($\omega = 0$).
- At $t = 0.2$ µs, the shock front reaches the initial location marked by the white circle in Figure 13a. This induces an instantaneous negative change of vorticity of $-10^6$ s$^{-1}$ and leads to the formation of a clockwise rotating vortex that matches the spin direction of the final recirculating pattern.
- Between $t = 0.2$ µs and 200 µs, the vorticity source terms $S_\mathrm{barocline}$, $S_\mathrm{viscous}$ and $S_\mathrm{expansion}$ progressively reduce the overall vorticity. Note that $S_\mathrm{stretching}$ is null in 2D and that $S_\mathrm{advection}$ is included in the material derivative.

These observations suggest that the vortices involved in the formation of the expelled torus are likely generated by the shock quasi-instantly near the tip of the electrode.

The analysis is taken a step further by computing the absolute value of the integrated sources of vorticity appearing on the right-hand side of Eqn. (13) along the trajectory of the vortex center. These integrated values are defined as:

$$I_\mathrm{mech}(t) \equiv \int_0^t S_\mathrm{mech}(t')dt' \quad [\mathrm{s}^{-1}] \tag{21}$$

The subscript "mech" stands for the mechanism inducing vorticity: "mech" is either "shock", "barocline", "viscous" or "expansion". The different $S_\mathrm{mech}$ are defined in Eqns. (14)–(20). The integral formalism used in Eqn. (21) was chosen to compare quantitatively on the same plot the effect of the scaled Dirac delta function $S_\mathrm{shock}$ and the other source terms. $I_\mathrm{expansion}$, $I_\mathrm{barocline}$ and $I_\mathrm{viscous}$ are computed by numerical integration of Eqn. (21), whereas $I_\mathrm{shock}$ is inferred from the vorticity jump predicted by the CFD method at 0.18 µs (Figure 14a). The results are shown in Figure 14b. The solid lines correspond to the sources that create negative vorticity ($I_\mathrm{mech}(t) < 0$, clockwise rotation), and the dashed lines to the sources of positive vorticity ($I_{\text{mech}}(t) > 0$, counterclockwise rotation).

Figure 14b demonstrates that the total amount of vorticity produced between 0 and 100 μs is mainly due to the sudden creation of vorticity by the shock at 0.2 μs (red curve), referred to as the ISIV mechanism $S_{\text{shock}}$. Then, this vorticity is slightly destroyed by the expansion mechanism $S_{\text{expansion}}$ (blue curve). The BT mechanism (green curve) contributes negligibly to the vorticity ($I_{\text{barocline}}/I_{\text{shock}} \approx 0.1$ at $t = 100$ μs). Furthermore, it produces vorticity in the opposite direction to the overall recirculation pattern which leads to the formation of the expelled torus.

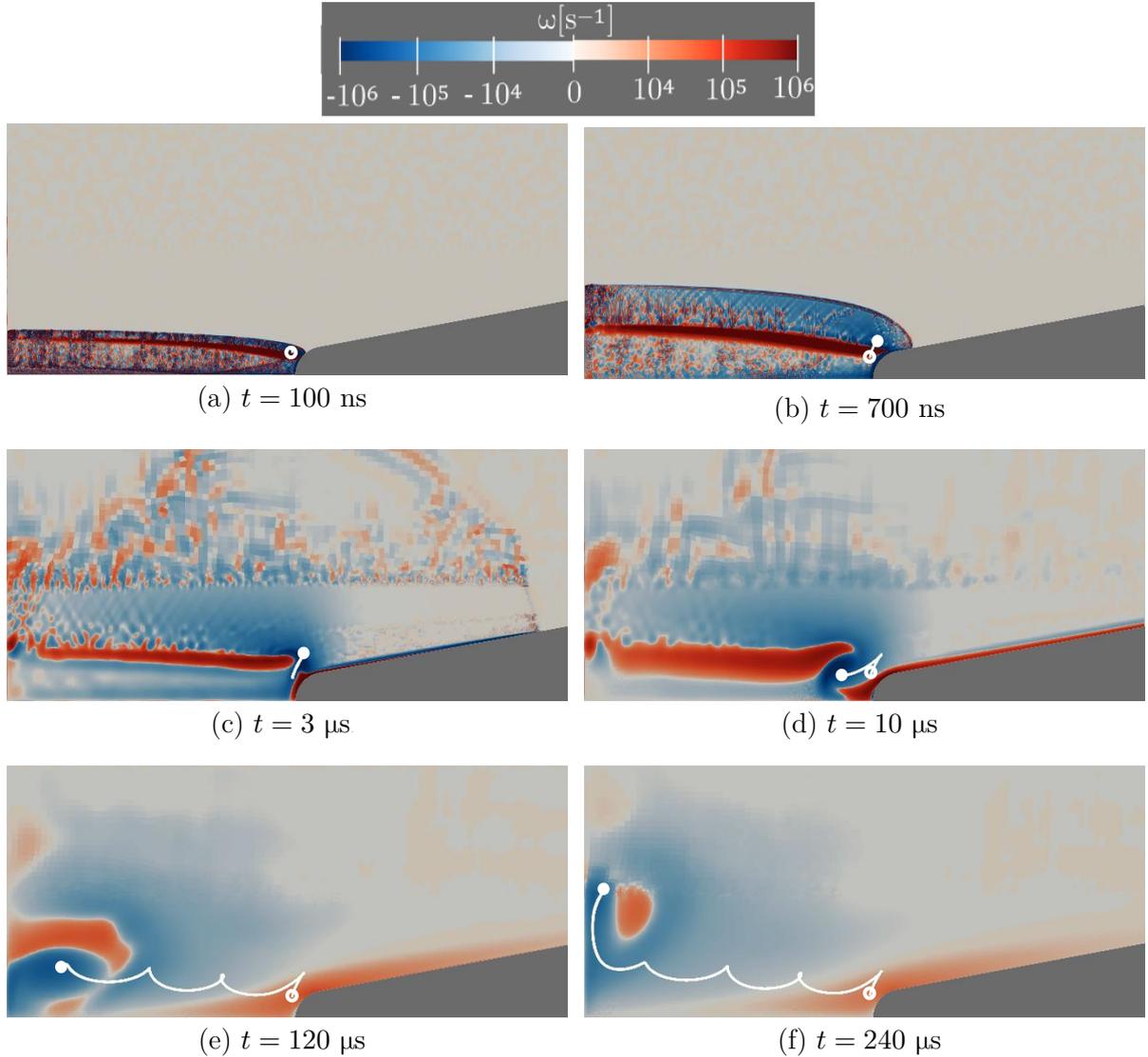

Figure 13. Computed vorticity fields for a discharge with a 3 mm gap length and a 6000 K initial plasma kernel temperature. The white line is the trajectory of a fluid element located initially near the electrode tip and then inside the core of the clockwise vortex. The white circle and the white dot are the initial and current locations of the fluid element, respectively.

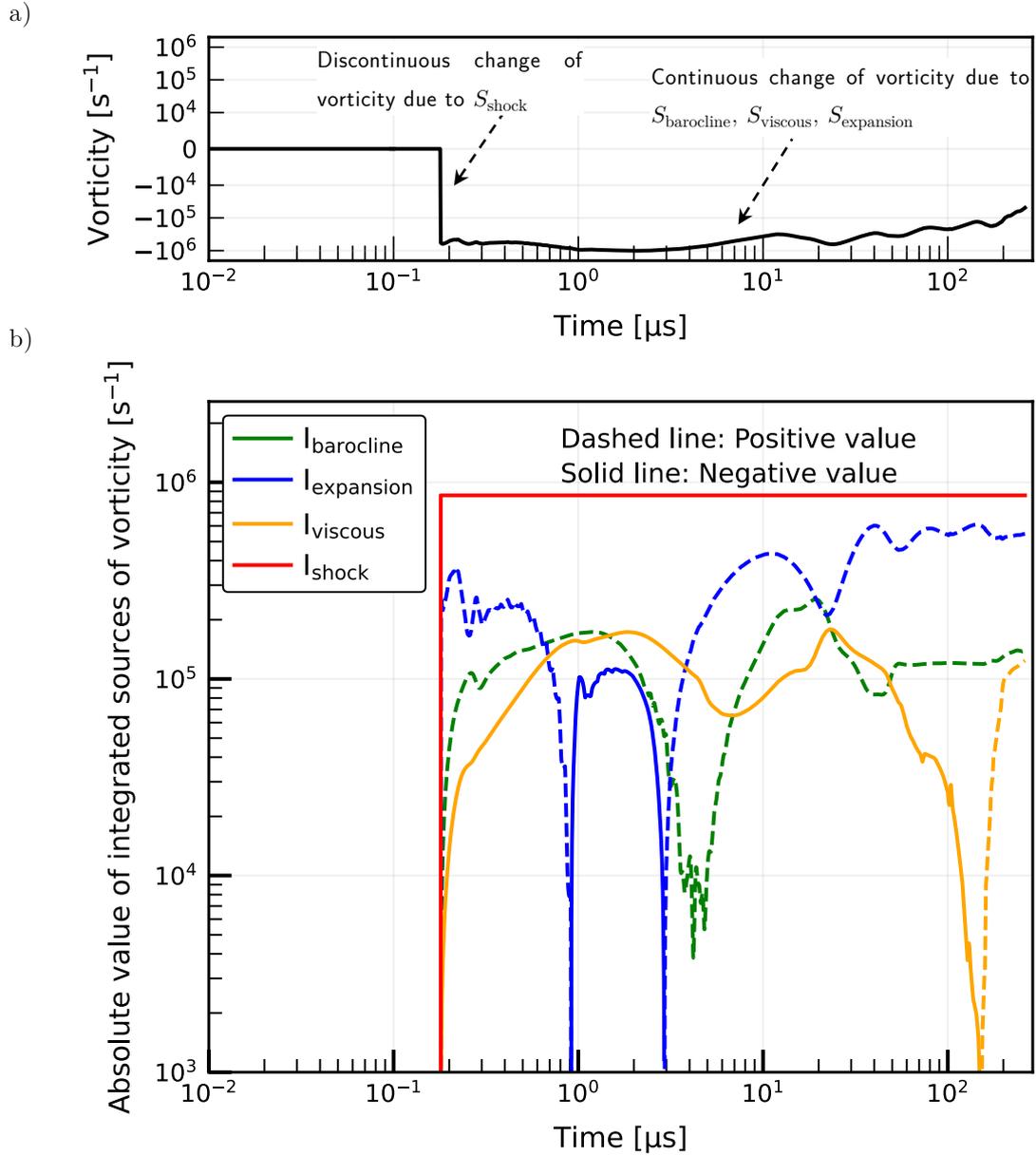

**Figure 14.** Vorticity (a) and absolute value of the integrated sources of vorticity (b) computed at the probe location for the discharge conditions: $T_k = 6000$ K, $L = 3$ mm. The vorticity sources correspond to the various terms on the RHS of Eqn. (13). Negative and positive integrated sources are represented by solid and dashed lines, respectively.

## IV.C. Impact of the temperature ratio on vorticity generation

The previous Subsection showed that the dominant vorticity source is the ISIV mechanism. We will now infer that the temperature ratio $T_k/T_{out}$ is the major factor driving the amount of vorticity created by this mechanism. Our demonstration, illustrated in Figure 15, is divided into three steps:

   i. First, Eqns. (19)–(20) state that, at any instant and any location along the shock front, the jump of vorticity $\Delta\omega$ is a function of the density ratio $\rho_1/\rho_0$ across the shock and of the shock normal velocity $u_{0,n}$. These two quantities are related to the local shock Mach number $M$ through the Rankine-Hugoniot relation and the definition $u_{0,n} = c_0 M$, respectively. Thus, **$\Delta\omega$ mostly depends on $M$**.
   ii. Second, for non-planar shock waves, Whitham [35] established that $M$ is related to the initial Mach number of the shock $M_s$ and some geometrical parameters accounting for the expansion of the shock front. Thus, **$M$ depends on $M_s$**.
   iii. Third, for any gas, $M_s$ is mainly governed by the temperature ratio $T_k/T_{out}$ (rather than by the individual temperatures $T_k$ and $T_{out}$ independently). To demonstrate this dependence, we make two key assumptions. On the one hand, we assume that the early-time evolution of the blast wave is analogous to that of an equivalent one-dimensional shock tube, as depicted in the right panel of Figure 15. Thus, the initial Mach number of the shock $M_s$ is related to the properties of the kernel and surrounding gases by the classical shock tube relations. Combining equations 7.13 and 7.94 from [34], we derive Eqn. (22).

$$\frac{P_{out}}{P_k}\left(\frac{2\gamma_{out}M_s^2}{\gamma_{out}+1} - \frac{\gamma_{out}-1}{\gamma_{out}+1}\right) = \left(1 - \frac{\gamma_k-1}{\gamma_{out}+1}(M_s - M_s^{-1})\frac{c_{out}}{c_k}\right)^{2\gamma_k/(\gamma_k-1)} \quad (22)$$

where $\gamma_k$ is the heat capacity ratio, $c_k = \sqrt{\gamma_k(R/\mathcal{M}_k)T_k}$ the sound speed and $\mathcal{M}_k$ the molar mass of the plasma kernel. The subscript "out" indicates that the quantities refer to the surrounding gas. On the other hand, we assume that the heating of the plasma kernel is isochoric during the pulse (because ultrafast heating occurs on a timescale much shorter than the expansion timescale $\tau_{rarefaction}$ [9,10]). The ideal gas law and the principle of mass conservation in a constant volume heat bath result in:

$$\frac{P_{out}}{P_k} = \frac{\rho_{out}(R/\mathcal{M}_{out})T_{out}}{\rho_k(R/\mathcal{M}_k)T_k} = \frac{\mathcal{M}_k}{\mathcal{M}_{out}}\frac{T_{out}}{T_k} \quad (23)$$

Injecting Eqn. (23) and the definition of sound speed into Eqn. (22), we obtain:

$$\frac{\mathcal{M}_k}{\mathcal{M}_{out}} \frac{T_{out}}{T_k} \left( \frac{2\gamma_{out} M_s^2}{\gamma_{out}+1} - \frac{\gamma_{out}-1}{\gamma_{out}+1} \right)$$
$$= \left( 1 - \frac{\gamma_k - 1}{\gamma_{out}+1}(M_s - M_s^{-1})\sqrt{\frac{\gamma_{out}\mathcal{M}_k T_{out}}{\gamma_k \mathcal{M}_{out} T_k}} \right)^{2\gamma_k/(\gamma_k-1)} \quad (24)$$

Eqn. (24) shows that $M_s$ is an implicit function (called $\mathcal{F}$) of the two adiabatic indices, the ratio of molar masses and the ratio of temperatures: $M_s = \mathcal{F}(T_k/T_{out}, \gamma_k, \mathcal{M}_k/\mathcal{M}_{out}, \gamma_{out})$. In Appendix B, we show that, in typical discharge conditions, the dependencies of $\mathcal{F}$ on $\gamma_{out}$, $\gamma_k$ and $\mathcal{M}_k/\mathcal{M}_{out}$ are small compared to that on $T_k/T_{out}$. Therefore, to first order, **$M_s$ is an implicit function of $T_k/T_{out}$**.

In conclusion, the vorticity induced by the ISIV mechanism $\Delta\omega$ is primarily governed by the ratio $T_k/T_{out}$.

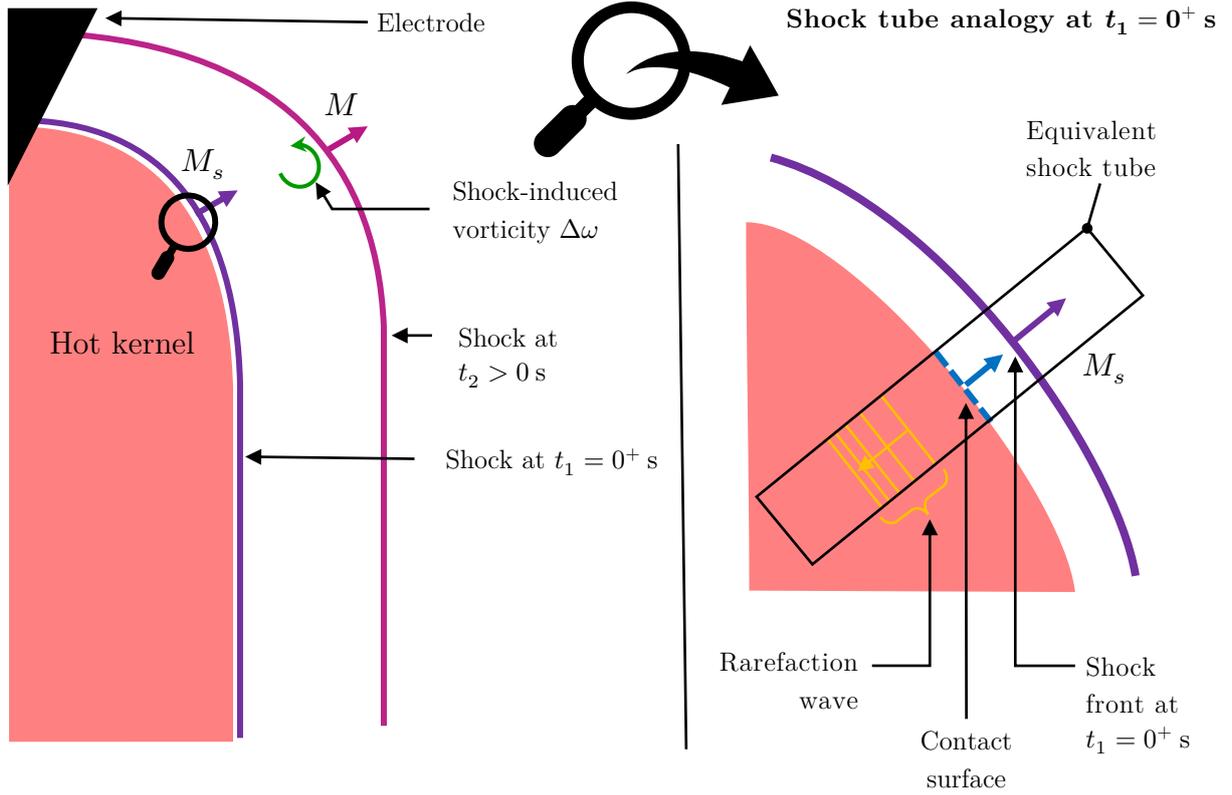

**Figure 15.** Schematic representation of the shock front expansion inducing vorticity $\Delta\omega$ between two instants, $t_1 = 0^+$ s and $t_2 > 0$ s. The violet curve depicts the initial shock front ($t_1$) with uniform Mach number $M_s$ along its profile. The magenta curve shows the evolved shock front ($t_2$) with non-uniform Mach number $M$. The right panel presents a magnified view of the left panel at $t_1$ and illustrates the shock tube analogy.

## V. A criterion for post-discharge flow dynamic regime

In this Section, we derive a new non-dimensional number characterizing the post-discharge flow dynamic regime (recirculating or not) based on the main findings of the present study. Then, we develop and discuss the criterion in light of results from the literature.

### V.A. Derivation of the characteristic number $\Pi^*$

As discussed in Subsection IV.B, our quantitative study of the vorticity sources showed that the collapse of the plasma kernel into an expelled torus is driven by the ISIV mechanism and not by the BT mechanism mentioned in [13–15].

Kono *et al.* [36] already observed that the recirculation is favored by high kernel temperatures. Other works, such as [15,37,38], emphasized the influence of the energy deposited by the discharges. Our study qualitatively agrees with these observations. However, the present work goes one step further as it shows that the vorticity produced by the ISIV mechanism depends primarily on $T_\mathrm{k}/T_\mathrm{out}$, rather than just on $T_\mathrm{k}$ or the energy input (see Subsection IV.C).

We therefore introduce a new dimensionless number, denoted as $\Pi^*$, to characterize the flow dynamic regime:

$$\Pi^* \equiv \frac{T_\mathrm{k}}{T_\mathrm{out}} \qquad (25)$$

$\Pi^*$ is an alternative to the dimensionless number $\Pi$ defined by Dumitrache *et al.* [15]:

$$\Pi \equiv \frac{E_\mathrm{uh}}{P_\mathrm{out} V} \qquad (26)$$

where $E_\mathrm{uh}$ is the energy deposited by the ultrafast heating processes induced by the discharge, $P_\mathrm{out}$ the ambient pressure, and $V$ the discharge volume.

$\Pi^*$ offers several advantages over $\Pi$ as a flow characterization:

- the vorticity created by the ISIV mechanism scales as $\Pi^*$ (as shown in Section IV.C). Hence, $\Pi^*$ has a physical interpretation, unlike $\Pi$, which was derived from a dimensional analysis and the Vaschy-Buckingham $\pi$ theorem [15].
- $\Pi^*$ is independent of the discharge radius (which is rarely precisely known).
- $\Pi^*$ is independent of the type of gas, contrary to $\Pi$. Indeed, let us consider two discharges, A and B, respectively in argon and methane, with the same ultrafast heating energies $E_\mathrm{uh}$ and the same $\Pi$ numbers. Due to the difference in the trans-rotational degrees of freedom (3 for A and 6 for B), discharge A heats up the gas twice as much as discharge B. Their $\Pi^*$ numbers will differ significantly, and as a consequence, the vortices magnitudes and the recirculating flows will be different. Unlike $\Pi^*$, the $\Pi$ number cannot capture this difference.

Hence, $\Pi^*$ appears as a natural identifier of the flow dynamic regime. However, we note that $\Pi^*$ may be more difficult to determine than $\Pi$ as it requires a quantitative spectroscopic measurement of the gas temperature at the end of the pulse.

It is interesting to note that $\Pi^*$ and $\Pi$ can be related by a simple analytical formula if the molar mass is constant during the pulse (i.e. with no significant dissociation or ionization) and if the energy going into ultrafast heating during the pulse can be expressed as $E_{\text{uh}} = V\rho_{\text{out}} c_v^{\text{tr-rot}}(T_k - T_{\text{out}})$. These two assumptions are usually valid for nanosecond discharges [9], except for thermal sparks. For a gas with $n_{\text{freedom}}$ trans-rotational degrees of freedom, the specific heat capacity determining the amount of heat going into ultrafast heating can be expressed as $c_v^{\text{tr-rot}} = (n_{\text{freedom}}/2) \times (R/\mathcal{M})$. In addition, the pressure is given by the ideal gas law, $P_{\text{out}} = \rho_{\text{out}} R T_{\text{out}}/\mathcal{M}$. Injecting these expressions into Eqn. (26), we get the following relation between $\Pi$ and $\Pi^*$:

$$\Pi = \frac{V\rho_{\text{out}}(n_{\text{freedom}}/2)R\mathcal{M}(T_k - T_{\text{out}})}{\rho_{\text{out}} R \mathcal{M} T_{\text{out}} V} = \frac{n_{\text{freedom}}}{2}\left(\frac{T_k}{T_{\text{out}}} - 1\right) \qquad (27)$$

Or equivalently:

$$\Pi^* = \frac{2}{n_{\text{freedom}}}\Pi + 1 \qquad (28)$$

### V.B. Flow regime criterion and comparison with the literature

Following the methodology of Dumitrache *et al.* [15], we plot in Figure 16 the dimensionless number $\Pi^*$ as a function of the interelectrode gap distance for several experimental [14,39–42] and simulated cases [14–17,36,43,44]. The cases reported in the Figure are those for which both the initial kernel temperature and the flow pattern are known.

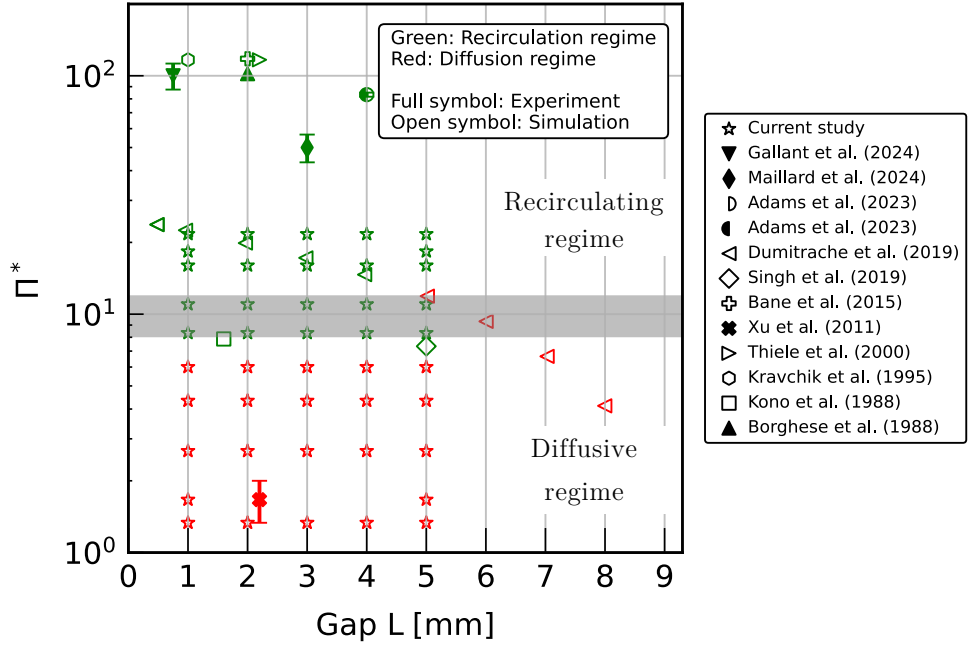

**Figure 16.** Dimensionless number $\Pi^* = T_k/T_{\text{out}}$ predicting the transition between the recirculating and diffusive regimes. The grayed area ($\Pi^* = 8\text{–}12$) represents the uncertainty in the transition based on a comparison between the CFD simulations from this study, as well as CFD simulations and experimental results from the literature [14–17,36,39–44]. Post-discharges in the recirculating and diffusive regimes are represented by green and red symbols, respectively.

The cases with a $\Pi^*$ number above the threshold 8–12 present a recirculation regime. Below this threshold, no recirculation pattern is reported, and the discharges are in the diffusive regime. Hence, $\Pi^*$ can be used as a convenient criterion to estimate the flow dynamic regime induced by nanosecond discharges. Essentially, this criterion compares the kinetic energy of the vortex ring with the energy dissipated by viscosity, which reduces its self-advection.

However, some limitations should be noted. First, most experiments reported in Figure 16 were conducted with single nanosecond pulses and in quiescent flows. When a pulse is preceded by a series of nanosecond pulses, the gas temperature in the gap may increase above $T_{\text{out}}$. In that case, $\Pi^*$ must be estimated with the actual temperature before the pulse instead of $T_{\text{out}}$. Second, if an external flow is imposed, the shock-induced vortices may be convected by the flow and the recirculation pattern may be different from the one studied in this article. Third, even though $\Pi^*$ characterizes the amount of vorticity created independently of the type of gas, the criterion threshold value (8–12) is not, as it is linked to the gas viscosity. Fourth, a recirculating flow does not necessarily mean that the core will return to ambient conditions, as the gap length and the pulse frequency also play a role and should not be too high for efficient cooling (see Figure 11).

## VI. Conclusions

In this study, we performed two-dimensional CFD simulations of the interpulse flow following a single nanosecond discharge. Depending on the temperature ratio, the gap length, and the frequency, we observed different flows and different cooling regimes. Focusing on a case with recirculation (representative of those observed in experiments), we identified the strongest vortex driving the main recirculation pattern and responsible for expelling the hot gases outside the interelectrode region. The vortex originates from a region close to the tip of the electrode. This region is crossed by the leading shock, whose strength is non-uniform owing to the shock's non-uniform curvature at that point. We computed all the vorticity sources along a trajectory of a fluid element belonging to the central vortex and compared their magnitude quantitatively. We found that the dominant mechanism responsible for the creation of the principal vortex is the quasi-instantaneous jump of vorticity induced by the non-uniform leading shock, which we refer to as the "Instantaneous Shock-Induced Vorticity" (ISIV) mechanism. On the time scale of the interpulse, the vorticity created by the ISIV mechanism is one order-of-magnitude higher than that created by the Barocline Torque (BT) mechanism. Moreover, the electrode shape plays a minor role in recirculation cooling, except if the electrodes are parallel walls. Finally, we demonstrated that the produced vorticity (and consequently the flow regime) is primarily governed by the non-dimensional number $\Pi^* = T_\mathrm{k}/T_\mathrm{out}$. Our CFD simulations indicate that discharges characterized by $\Pi^* \gtrsim 8$–$12$ exhibit a recirculation regime, and those below that threshold are in the diffusive regime. This finding is supported by comparisons with simulations and experiments from the literature.

## Acknowledgements

The authors are grateful to Prof. Aymeric Vié and Dr. Christian Tenaud for valuable discussions on the topic. This work has received funding from the Ministerial Allocation for Polytechnique students (AMX) and the European Research Council (ERC) under the European Union's Horizon 2020 research and innovation program (ERC GreenBlue grant agreement No.101021538). A CC-BY 4.0 public copyright license has been applied by the authors to the present document and will be applied to all subsequent versions up to the Author Accepted Manuscript arising from this submission, in accordance with the grant's open access conditions: https://creativeco mmons.org/licenses/by/4.0/.

# Appendices

## A Sensitivity of the CFD model to the assumptions on discharge and simulation parameters

We tested the sensitivity of the mean interelectrode temperature at the end of the interpulse $\overline{T}_{\text{end}}$ to the following discharge and simulation parameters:

- the presence or absence of electrodes,
- the shape of the electrodes,
- the initial shape of the plasma kernel,
- the initial radius of the plasma kernel,
- the steepness of the transition at the edge of the kernel shape while initializing the thermodynamic properties (pressure, temperature),
- the presence of preceding pulses,
- the chemistry model.

Thirteen new CFD simulations were conducted, changing the parameters one by one from the reference case corresponding to an elliptical kernel, a pin electrode, an infinitely fast chemistry (LTE), no previous pulse, $R_{\text{d}} = 100$ µm, and the following configuration: $L = 3$ mm, $T_{\text{k}} = 6000$ K. Figure 17 shows the various kernel profiles considered, with their different shapes or radii. Figure 18 illustrates the different electrode shapes tested (including a case with no electrode). We also tested two other chemistry models to study the impact of the LTE assumption:

- First, we assumed that the chemistry is frozen, meaning that thermodynamics and transport properties are kept constant at their initial values.
- Second, we used the finite rate chemistry model developed by Park [45].

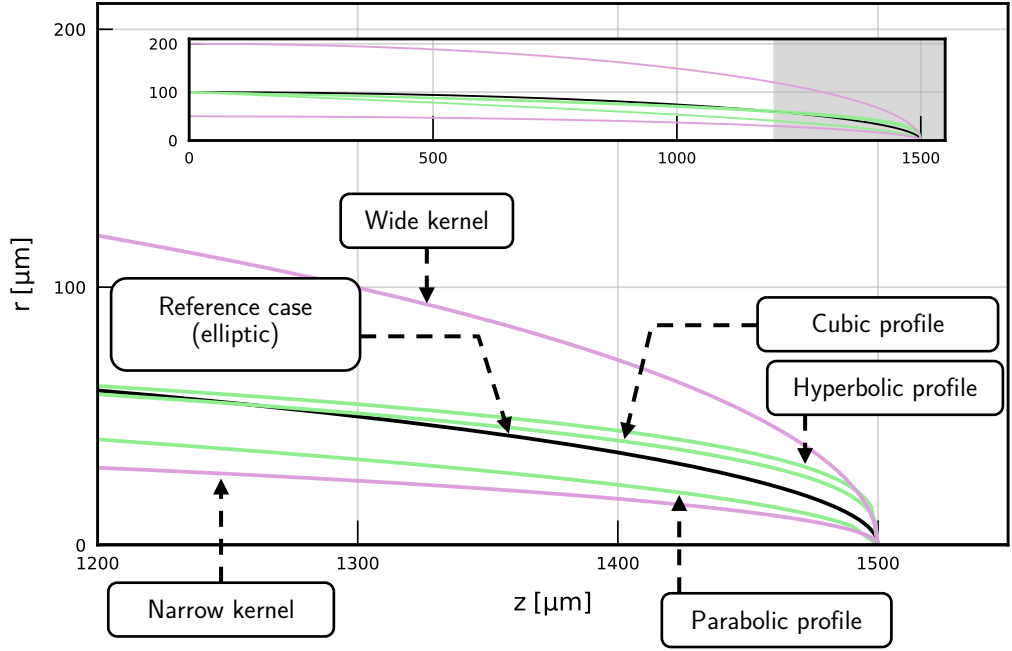

figure 17. Shapes of the plasma kernel used to initialize the CFD simulations. Shapes with various profiles (elliptic, cubic, hyperbolic, parabolic) are shown in green. Elliptic shapes with various radii (50, 100 and 200 µm) are shown in purple. The main graph zooms in on the near-electrode region. The full profiles are shown in the inset.

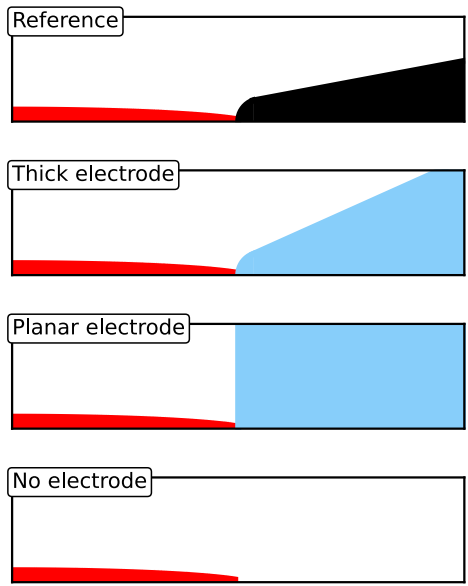

Figure 18. Electrode shapes tested in CFD simulations. Shapes with varying profiles (thick, planar) are shown in blue. The last case does not compound any electrode.

Figure 19 compares the simulated mean temperature $\overline{T}_\text{end}$ at the end of the post-discharge with that of the baseline simulation at $f = 10$ kHz. In all cases (except the planar electrode), $\overline{T}_\text{end}$ is between 300 and 600 K, hence much lower than $T_\text{k}$. Thus, to first order, the post-discharge temperature $\overline{T}_{\text{end}}$ is not sensitive to the discharge parameters (as long as electrodes are pin-like). In the case of planar electrodes, cooling is less efficient although $\overline{T}_{\text{end}}$ is still much lower than $T_{\text{k}}$. We identified two explanations:

- the electrode walls prevent the initial elliptical uniform shock from fully transforming into a spherical shock, thus reducing the non-uniform shock development and the strength of the ISIV mechanism,
- the convection of cold gases into the recirculating torus is hindered by the presence of the walls.

Figure 19 also shows that the three chemistry models (LTE, frozen chemistry, finite rate chemistry) produce similar cooling profiles and flow patterns, resulting in similar values of $\overline{T}_{\text{end}}$. We chose to use the LTE model in the rest of the article because the equilibrium thermodynamic and transport properties of air are well-documented.

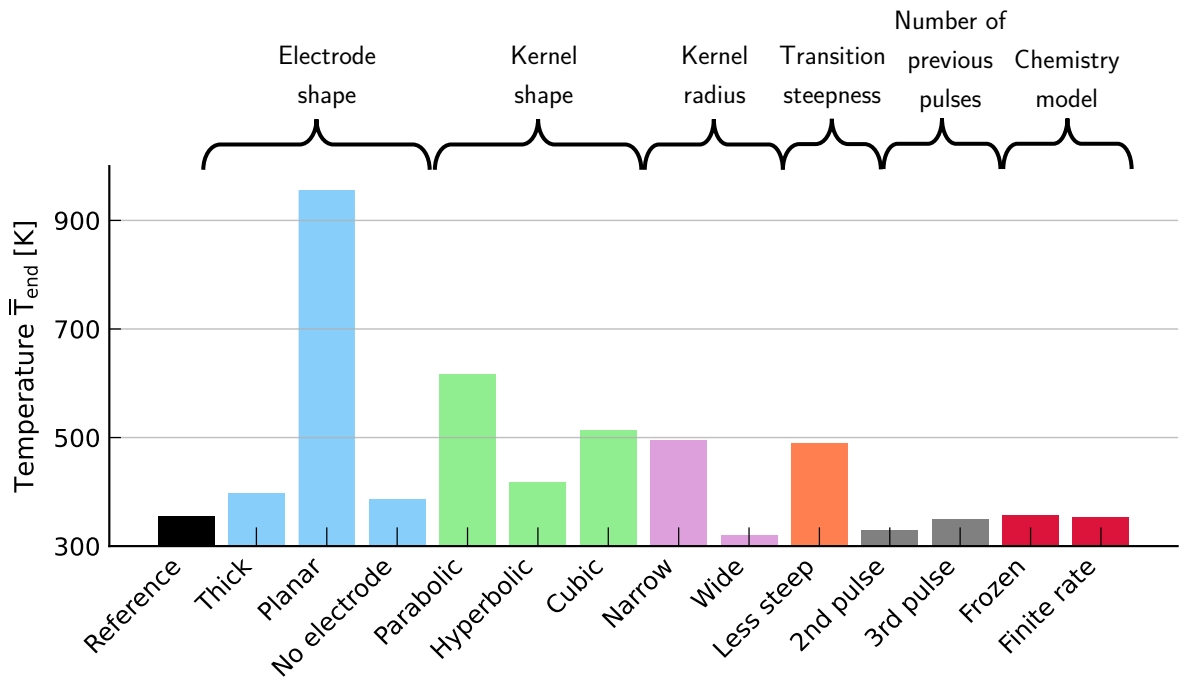

**Figure 19.** Mean temperature predicted by CFD simulations at the end of the interpulse for different discharge and simulation parameters (electrode shape, kernel shape, steepness of the transition between the kernel and the ambient properties, discharge radius, existence of previous pulses, chemistry model). The temperature is averaged along the pin-to-pin axis. The discharge configuration corresponds to $L = 3$ mm, $T_{\text{k}} = 6000$ K with a frequency of 10 kHz.

## B  Sensitivity of the initial Mach number to $T_{\text{k}}/T_{\text{out}}$ and $T_{\text{k}}$

In Section IV.C, we showed that $M_s$ depends implicitly on $T_{\text{k}}/T_{\text{out}}$, $\gamma_{\text{k}}$, $\gamma_{\text{out}}$ and $\mathcal{M}_{\text{k}}/\mathcal{M}_{\text{out}}$ through Eqn. (24). In this Appendix, we demonstrate that, in typical nanosecond discharges, the dependence of $M_s$ on $\gamma_\text{k}$, $\gamma_\text{out}$, and $\mathcal{M}_\text{k}/\mathcal{M}_\text{out}$ is mild compared to that on $T_\text{k}/T_\text{out}$.

To this end, we select five representative cases corresponding to typical nitrogen plasma discharges:

- Case 1: the translational and rotational modes of $N_2$ in the plasma kernel are fully excited, and the vibrational mode is assumed to be frozen. The ambient gas is pure $N_2$ and its vibrational mode is not excited.
- Case 2: the translational, rotational, and vibrational modes in the plasma kernel are fully excited. The vibrational mode of the ambient gas is not excited.
- Case 3: the translational, rotational, and vibrational modes in the plasma kernel are fully excited. The vibrational mode of the ambient gas is fully excited.
- Case 4: the gas in the plasma kernel is fully dissociated ($N_2 \to N + N$). The vibrational mode of the ambient gas is not excited.
- Case 5: the gas in the plasma kernel is fully dissociated and fully ionized ($N_2 \to N + N \to 2N^+ + 2e^-$). The vibrational mode of the ambient gas is not excited.

Each case corresponds to a typical discharge regime: glow discharges [6] and non-thermal sparks [7] (Cases 1–4) or thermal sparks [8] (Case 5). Table 1 lists the corresponding values of heat capacity ratios ($\gamma_\text{k}$, $\gamma_\text{out}$) and molar masses ($\mathcal{M}_\text{k}$, $\mathcal{M}_\text{out}$) for the plasma kernel and the ambient gas.

We numerically solved Eqn. (24) for each case. The resulting values of $M_s$ are shown in Figure 20. We see that $M_s$ is governed primarily by $T_\text{k}/T_\text{out}$, whereas the dependence on $\gamma_\text{k}$, $\gamma_\text{out}$ and $\mathcal{M}_\text{k}/\mathcal{M}_\text{out}$ is mild: increasing $T_\text{k}/T_\text{out}$ by a factor 100 induces a factor ~10 increase on $M_s$, whereas changing the conditions of the plasma kernel from Cases 1 to 5 increases $M_s$ by less than 50%. The curves for Cases 1–3 overlap within 3% and are effectively represented by a single curve (red). This indicates that the degree of excitation of the vibrational mode of the plasma kernel or of the ambient gas has a negligible influence on $M_s$.

Table 1. Thermodynamic parameters of the plasma kernel and ambient gas for five typical nitrogen plasma discharges.

| N° | Note | $\gamma_{\text{out}}$ | $\mathcal{M}_{\text{out}}$ [g/mol] | $\gamma_{\text{k}}$ | $\mathcal{M}_{\text{k}}$ [g/mol] |
|---|---|---|---|---|---|
| 1 | Plasma kernel: frozen vibration<br>Ambient gas: frozen vibration | 7/5 | 28 | 7/5 | 28 |
| 2 | Plasma kernel: excited vibration<br>Ambient gas: frozen vibration | 7/5 | 28 | 9/7 | 28 |
| 3 | Plasma kernel: excited vibration<br>Ambient gas: excited vibration | 9/7 | 28 | 9/7 | 28 |
| 4 | Plasma kernel: fully dissociated<br>Ambient gas: frozen vibration | 7/5 | 28 | 5/3 | 14 |
| 5 | Plasma kernel: fully dissociated and fully ionized<br>Ambient gas: frozen vibration | 7/5 | 28 | 5/3 | 7 |

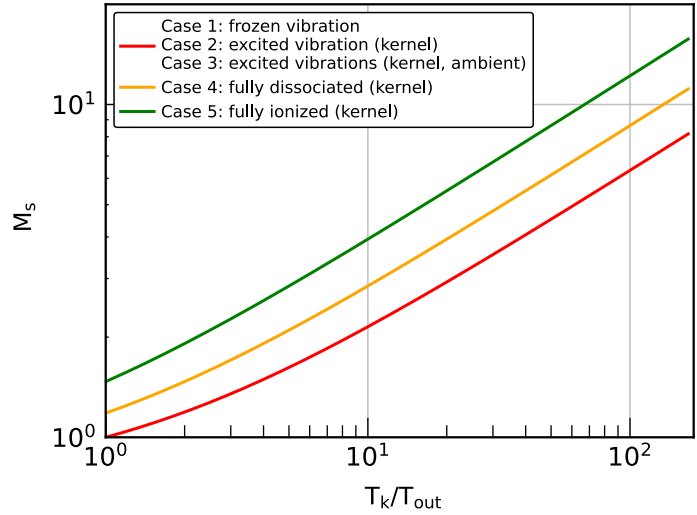

Figure 20. Evolution of the initial Mach number $M_s$ of the leading shock as a function of $T_{\text{k}}/T_{\text{out}}$. Values of $M_s$ are calculated by solving Eqn. (24) for the five distinct discharge conditions ($\gamma_{\text{k}}, \mathcal{M}_{\text{k}}, \gamma_{\text{out}}, \mathcal{M}_{\text{out}}$) specified in Table 1. Results for Cases 1 to 3 overlap and are represented by a single red curve.